\documentclass[12pt]{article}
\usepackage[T2A]{fontenc}
\usepackage[english]{babel}
\def\frak#1{\cal #1}
\def\un{{\rm 1\mkern-4mu I}}
\usepackage{amssymb,amsmath}
\usepackage{graphicx}
\usepackage{alltt}
\usepackage{enumerate}
\newtheorem{theorem}{Theorem}%[section]
\begin{document}
\begin{titlepage}		
	\vspace{1in}
	
	\begin{center}
{\Large{\bf{ Octonions, exceptional Jordan algebra,
			and the role of the group F4 in particle physics\footnote{Extended version of lectures presented by I.T. at the Institute for Nuclear Research and Nuclear Energy during the spring of 2017 (see \cite{TD17}) and in May 2018 (published in:
\textit{Adv. Appl. Clifford Alg.} \textbf{28} (2018) 82).}}}
	\vspace{4mm}
	
	\large{Ivan Todorov and Svetla Drenska} \\[10pt]
	\small
	Institute for Nuclear Research and Nuclear Energy \\
    Bulgarian Academy of Sciences \\
	\small    Tsarigradsko Chaussee 72, BG-1784 Sofia, Bulgaria\\	
	\small  e-mail: ivbortodorov@gmail.com, \, sdren@abv.bg}

\end{center}

\begin{abstract}
Normed division rings are reviewed in the more general framework of composition algebras that include the split (indefinite metric) case. The Jordan - von Neumann - Wigner classification of finite euclidean Jordan algebras is outlined with special attention to the 27 dimensional exceptional Jordan algebra $\mathcal{J}$. The automorphism group $F_4$ of $\mathcal{J}$ and its maximal Borel - de Siebenthal subgroups $\frac{SU(3)\times SU(3)}{{\mathbb{Z}_3}}$ and Spin(9) are studied in detail and applied to the classification of fundamental fermions and gauge bosons. Their intersection in $F_4$ is demonstrated to coincide with the gauge group of the Standard Model of particle physics. The first generation's fundamental fermions form a basis of primitive idempotents in the euclidean extension of the Jordan subalgebra $J_2^8$ of $\mathcal{J}$.\\

\bigskip

	
\end{abstract}	

%\vfill\eject
\end{titlepage}

\tableofcontents

\setcounter{equation}{0}
\renewcommand\theequation{0.\arabic{equation}}

\section*{Introduction}	
\setcounter{equation}{0}
\renewcommand\theequation{\thesection.\arabic{equation}}

Division and Clifford algebras were introduced in 19th century with an eye for applications in geometry and physics (for a  historical  survey see the last chapter  of \cite{L}). Pascual Jordan introduced and studied his algebras in the 1930's
in order to describe observables in quantum mechanics (for "a taste of Jordan algebras" see \cite{McC} along with the original paper \cite{JvNW}). Yet, the first serious applications of these somewhat exotic structures appeared (in mid-twentieth century) in pure mathematics: in the theory of exceptional Lie groups and symmetric spaces (cf. \cite{F} as well as the later surveys \cite{A,K, BS,R08}), in topology \cite{ABS}. For an entertaining review on division algebras and quantum theory - see \cite{B12}. Possible applications to particle physics were first advocated by Feza G\"{u}rsey and his students in the 1970's - see his lecture and his posthumous book (with C.-H.Tze) \cite{G} and references therein). They continue in various guises to attract attention until these days, never becoming a mainstream activity. The present lectures  are meant as a background for the ongoing work \cite{DV, TD, D-VT} centered on the exceptional 27 dimensional Jordan algebra $\mathcal{J} =
\mathcal{H}_3(\mathbb{O})$. In Chapter 4 we elaborate on a possible application of the automorphism group $F_4$ of
$\mathcal{J}=J_3^8$  as a novel "grand unified symmetry" of the Standard Model. In Chapter 5 we apply the notion of extended special euclidean Jordan algebra of \cite{D-VT} to the spin factor $J_2^8$ and  put forward a fermion Fock space construction, inspired by \cite{F18}, of the first generation of leptons and quarks. Although such a proposal of an "exceptional finite quantum geometry" is still tentative, we feel that it is worth pursuing.\footnote{For related attempts to provide an algebraic counterpart of the Standard Model of particle physics see \cite{D04, D10, D14, CGKK, F16, S18, F18} and references therein.} In any case, the mathematical background which is the main subject of these notes is sound and beautiful - and deserves to be known by particle theorists.

\section{Composition and Clifford algebras}
\subsection{Normed alternative algebras (\cite{R08} Sect.1)}
A composition (or Hurwitz) algebra ${\mathcal A}$ is a vector space over a field $\mathbb{K}=(\mathbb{R},\mathbb{C},\dots)$ equipped with a bilinear (not necessarily associative) product $xy$ with a unit $1$ ($1 x=x1=x$) and a nondegenarate quadratic form $N(x)$, the \textit{norm}, satisfying
\begin{equation}
N(xy)=N(x)N(y), \quad  N(\lambda x) =\lambda^2 N(x) \,\ \text{for} \,\ x \in {\mathcal A}, \ \lambda\in \mathbb{K}.
\label{1.1}
\end{equation}
The norm allows to define by polarization a \textit{symmetric bilinear form} $<x,y>$ setting:
\begin{eqnarray}
2<x,y>=N(x+y)-N(x)-N(y)(=2<y,x>).
\label{1.2}
\end{eqnarray}
(Nondegeneracy of $N$ means that if $<x,y>=0$ for all $y \in {\mathcal A}$ then $x=0$.)
By repeated polarization of the identity $<xy,xy>=<x,x><y,y>$ one obtains
\begin{eqnarray}
<ab,ac>=N(a)<b,c>=<ba,ca>
\label{1.3}
\end{eqnarray}
\begin{eqnarray}
<ac,bd>+<ad,bc>=2<a,b><c,d>.
\label{1.4}
\end{eqnarray}
Setting in (\ref{1.4}) $a=c=x$, $b=1$, $d=y$ and using (\ref{1.3}) we find:
$$ <x^2+N(x)1-t(x)x,y> =0,  $$
where $t(x):= 2<x,1>$ is by definition the \textit{trace}, or, using the non-degeneracy of the form $< , >$,
\begin{eqnarray}
x^2-t(x)x+N(x)1=0,\qquad t(x)=2<x,1>.
\label{1.5}
\end{eqnarray}
Thus every $x \in {\mathcal A}$ satisfies a quadratic relation with coefficients  the trace $t(x)$ and the norm $N(x)$ (a linear and a quadratic scalar functions) taking values in $\mathbb{K}$.

The trace functional (\ref{1.5}) allows to introduce \textit{Cayley conjugation},
\begin{eqnarray}
x \rightarrow x^*=t(x)-x, \qquad t(x)=t(x)1\in {\mathcal A}
\label{1.6}
\end{eqnarray}
an important tool in the study of composition algebras. It is an (orthogonal) reflection $(<x^*,y^*>=<x,y>)$ that leaves the scalars $\mathbb{K}1$ invariant (in fact, $t(\lambda 1)=2\lambda$ implying $(\lambda 1)^*= \lambda 1$ for $\lambda\in\mathbb{K}$). It is also an involution and an antihomomorphism:
\begin{eqnarray}
(x^*)^*=x, \qquad (xy)^*=y^*x^*.
\label{1.7}
\end{eqnarray}
Furthermore Eqs.(\ref{1.5}) and (\ref{1.6} ) allow to express the trace and the norm as a sum and a product of $x$ and $x^*$:
\begin{eqnarray}
t(x)=x+x^*,\qquad N(x)=xx^*=x^*x=N(x^*). \nonumber
\end{eqnarray}
The relation (\ref{1.4}) allows to deduce
\begin{eqnarray}
<ax,y>=<x,a^*y>, \qquad <xa,y>=<x,ya^*>.
\label{1.8}
\end{eqnarray}
From these identities it follows $<ab,1>=<a,b^*>=<ba,1>$, hence, the trace is commutative:
\begin{eqnarray}
t(ab)=<b,a^*>=<a,b^*>=t(ba).
\label{1.9}
\end{eqnarray}
Similarly, one proves that $t$ is associative and symmetric under cyclic permutations.
\begin{eqnarray}
t((ab)c)=t(a(bc))=:t(abc)=t(cab)=t(bca).
\label{1.10}	
\end{eqnarray}
Moreover, using the quadratic relation (\ref{1.5}) and the above properties of the trace one proves the identities that define an \textit{alternative algebra}:	
\begin{eqnarray}
x^2y=x(xy),\qquad yx^2=(yx)x
\label{1.11}
\end{eqnarray}
(see Sect.1 of \cite{R08} for details). The conditions (\ref{1.11}) guarantee that the \textit{ associator}
\begin{eqnarray}
[x,y,z]=(xy)z-x(yz)
\label{1.12}
\end{eqnarray}
changes sign under odd permutations (and is hence preserved by even, cyclic, permutations). This implies, in particular, the \textit{flexibility conditions}.
\begin{eqnarray}
(xy)x=x(yx).
\label{1.13}
\end{eqnarray}

An unital alternative algebra with an involution $x\rightarrow x^*$ satisfying (\ref{1.7})  is a composition algebra if the norm $N$ and the trace $t$ defined by (\ref{1.9}) are scalars (i.e. belong to $\mathbb{K}(=\mathbb{K} 1)$) and the norm is non-degenerate.

Given a finite dimensional composition algebra ${\mathcal A}$ Cayley and Dickson have proposed a procedure to construct another composition algebra ${\mathcal A'}$ with twice the dimension of ${\mathcal A}$. Each element $x$ of ${\mathcal A'}$ is written in the form
\begin{eqnarray}
x=a+eb,\ a,\ b\in {\mathcal A}
\label{1.14}
\end{eqnarray}
where $e$ is a new "imaginary unit" such that
\begin{eqnarray}
e^2=-\mu,\ \mu \in \{ 1,-1\} \ (\mu^2=1).
\label{1.15}
\end{eqnarray}
Thus ${\mathcal A}$ appears as a subalgebra of ${\mathcal A'}$. The product of two elements $x=a+eb$, $y=c+ed$ of ${\mathcal A'}$ is defined as
\begin{eqnarray}
xy=ac-\mu d	\overline{b}+e(\overline{a}d+cb)
\label{1.16}
\end{eqnarray}
where $a \rightarrow \overline{a}$ is the Cayley congugation in ${\mathcal A}$. (The order of the factors becomes important, when the product in ${\mathcal A}$ is noncommutative.)
The Cayley conjugation $x\rightarrow x^*$ and the norm $N(x)$ in ${\mathcal A'}$ are defined by:
\begin{eqnarray}
x^*=(a+eb)^*=\overline{a}+\overline{b}e^*=\overline{a}-\overline{b}e=\overline{a}-eb \nonumber  \\
N(x)=xx^*=a\overline{a}+\mu b\overline{b}=x^*x.
\label{1.17}
\end{eqnarray}

Let us illustrate the meaning of (\ref{1.16}) and (\ref{1.17}) in the first simplest cases.

For ${\mathcal A}=\mathbb{R}$ , $\overline{a}=a$, Eq.(\ref{1.16}) coincides with the definition of complex numbers for $\mu=1$ ($e=i$) and defines the split complex numbers for $\mu=-1$. Taking next ${\mathcal A}=\mathbb{C}$ and $\mu=1$ we can identify ${\mathcal A'}$ with a $2\times2$ matrix representations setting
\begin{eqnarray}
\textbf{a}= a_0 +e_1a_1=a_0+i\sigma_3a_1= \left( \begin{array}{cc}
a & 0 \\
0 &\overline {a} \\
\end{array} \right) \ (a=a_0+ia_1) \nonumber \\
  \end{eqnarray}
\begin{eqnarray}
x= \textbf{a} +e\textbf{b},\ e=\left( \begin{array}{cc}
0 & -1 \\
1 &  0 \\
\end{array} \right) \Rightarrow x= \left( \begin{array}{cc}
a & -\overline{b}  \\
b &  \overline{a}  \\
\end{array} \right), \,
 \textbf{b}=b_0+e_1b_1. \nonumber
\end{eqnarray}
Anticipating Baez \textit{Fano plane} \cite{B} notation for the octonion imaginary units (see Appendix A) we shall set $e=e_4$,\quad $e_4e_1=e_2 \,\ (=i\sigma_1)$.

It is then easily checked that the multiplication law (\ref{1.16}) reproduces the standard matrix multiplication, the Cayley conjugation $x\rightarrow x^*$ coincides with the hermitian conjugation of matrices, while the norm $N(x)$ in ${\mathcal A'}$ is given by the determinant:
\begin{eqnarray}
\mathbb{H}=\{x\in \mathbb{C}[2]; \,\ xx^*=\det x(\geq 0)\}.
\label{1.19}
\end{eqnarray}

Similarly, starting with the split complex numbers, we can write
\begin{eqnarray}
\bf{a}_s=a_0+\tilde{e}_1a_1,\ \tilde{e}_1=\sigma_3\leftrightarrows  \textbf{a}_s= \left( \begin{array}{cc}
a_s & 0  \\
0 &  \overline{a}_s  \\
\end{array} \right) \ (a_s=a_0+a_1, \ \overline {a}_s=a_0-a_1).\nonumber
\end{eqnarray}
and choosing the same $e$ as above we can identify the \textit{split quaternions} $\mathbb{H}_s$ with real $2\times2$ matrices:
\begin{eqnarray}
\mathbb{H}_s=\{x= \left( \begin{array}{cc}
a_s & -\overline{b}_s  \\
b_s &  \overline{a}_s  \\
\end{array} \right) \ \in \mathbb{R}[2], \
  x^*=\left( \begin{array}{cc}
\overline{a}_s & \overline{b}_s  \\
-b_s &  a_s  \\
\end{array} \right),\ xx^*=\det x\}
\label{1.20}
\end{eqnarray}
its norm having signature $(2,2)$.

The next step in Cayley-Dickson construction gives the octonions, which have a nonassociative (but alternative) multiplication and thus do not have matrix realization.

\subsection{Relation to Clifford algebras. Classification.}
Given a composition algebra ${\mathcal A}$ we define subspace ${\mathcal A}_0 \subset {\mathcal A}$ of pure imaginary elements with respect to the Cayley conjugation (\ref{1.6}):
\begin{eqnarray}
{\mathcal A}_0=\{y \in{\mathcal A}; \, \,  y^*=-y \}.
\label{1.21}
\end{eqnarray}
It is a subspace of co-dimension one, orthogonal to the unit $1$ of ${\mathcal A}$ . For any $x\in{\mathcal A} $ we define its imaginary part as
\begin{eqnarray}
x_0=\frac{1}{2}(x-x^*)=x-<x,1>\Rrightarrow <x_0,1>=0.
\label{1.22}
\end{eqnarray}
From the expression $N(x)=xx^*$ (\ref{1.8}) and from the defining property (\ref{1.21}) of imaginary elements it follows that
\begin{eqnarray}
x_0 \in {\mathcal A}_0 \Rightarrow x_0^2=-N(x_0).
\label{1.23}
\end{eqnarray}
In other words, if the composition algebra ${\mathcal A}$ is $n$-dimensional then its $(n-1)$-dimensional subalgebra ${\mathcal A}_0$ gives rise to a Clifford algebra. If the norm $N$ is positive definite then \footnote{We adopt the sign convention of \cite{L}, \cite{G}, \cite{T}; \, the opposite sign convention, $C\ell_{(n-1)}$ for the positive definite $N(x)$, is used e.g. in \cite{B}.} ${\mathcal A}_0=C\ell(0,n-1)=C\ell_{(1-n)}$. In the case of \textit{split} complex numbers, quaternions and octonions one encounters instead the algebras $C\ell_1 \equiv C\ell(1,0),\ C\ell(2,1)$ and $C\ell(4,3)$, respectively.

It turns out that the classification of the Clifford algebras $C\ell_{(1-n)}$ implies the classification of normed division rings of dimension $n$. So we recall it in Table 1.
\begin{table}
%	\caption{{Irreducible spinors in the Clifford algebras $C\ell_{(1-n)}$.} 
%	\begin{center}
	\begin{tabular}{cccccc}
		\hline
	$n$ & $C\ell_{(1-n)}$ & Irreducible spinor &$n$ & $C\ell_{(1-n)}$ & Irreducible spinor \\
	\hline
	$1$ & $\mathbb{R}$ & $S_1=\mathbb{R}$ & $5$ & $\mathbb{H}[2]$ & $S_5= \mathbb{H}^2$ \\
	$2$ & $\mathbb{C}$ & $S_2=\mathbb{C}$ & $6$ & $\mathbb{C}[4]$ & $S_6= \mathbb{C}^4$ \\
	$3$ & $\mathbb{H}$ & $S_3=\mathbb{H}$ & $7$ & $\mathbb{R}[8]$ & $S_7= \mathbb{R}^8$ \\
	$4$ & $\mathbb{H} \oplus \mathbb{H}$ & $S_4^+=\mathbb{H}$,\ $S_4^-=\mathbb{H}$  & $8$ & $\mathbb{R}[8] \oplus \mathbb{R}[8]$ & $S_8^+= \mathbb{R}^8$,\ $S_8^-= \mathbb{R}^8$ \\
\hline
\end{tabular}
	\caption{Irreducible spinors in the Clifford algebras $C\ell_{(1-n)}$.}
%\end{center}
\end{table}
Here we use the notation $\mathbb{A} [n]$ for the algebra of $n \times n$ matrices with entries in the (associative)
algebra $\mathbb{A}$. As discovered by \'Elie Cartan in 1908  $C\ell_{(-\nu-8)} = C\ell_{-\nu} \otimes \mathbb{R}[16]$ so that the above Table 1 suffices to reconstruct all Clifford algebras  of type $C\ell_{-\nu}$. We see that the (real) dimension of the irreducible representation of $C\ell_{(1-n)}$ coincides with $n$ for $n=1, 2, 4, 8$ only, thus implying Hurwitz theorem (see
\cite{B} Theorem 1 and the subsequent discussion).

Proceeding to the split alternative composition algebras we note that the type of $C\ell(p,q)$ only depends on the signature $p-q$ which is $1$ (similar to $-7$ modulo $8$); we have: $C\ell(1,0)=\mathbb{R} \oplus \mathbb{R}$, $C\ell(2,1)=\mathbb{R}[2] \oplus \mathbb{R}[2]$, $C\ell(4,3)=\mathbb{R}[8] \oplus \mathbb{R}[8]$, for all above cases
\begin{eqnarray}
C\ell(p,p-1)\cong \mathbb{R}[2^{(p-1)}] \oplus \mathbb{R}[2^{(p-1)}].
\label{1.24}
\end{eqnarray}
We note here the difference in the treatment of the representations of $C\ell(p,p-1)$ in the cases $p=1,2$, in which we are dealing with real associative composition algebras $\mathbb{C}_s$ and $\mathbb{H}_s$, and $p=4$ of the split octonions. In the associative case we deal with the action of $C\ell(p,p-1)$ on the direct sum $\mathbb{R}^n \oplus \mathbb{R}^n$, $n=2^{(p-1)}$ (for $p=1,2$) while in the non-associative case it acts on the irreducible subspace $\mathbb{R}^n$ ($n=8$), thus again fitting the dimension of the corresponding alternative algebra.

\textit{Remark} 1.1. The \textit{spinors} $S_n$ are  here understood as quantities transforming under the lowest order faithful irreducible representation of the (compact) group Spin(n)  which consists of the norm one even elements of $C\ell_{-n}$. In fact, the even part $C\ell_0(p,q)$ of $C\ell(p,q)$ is isomorphic, for $q>0$, to $C\ell(p,q-1)$. Spin(n) is the double cover of the rotation group $\rm{SO}(n)$. The group of all norm one elements $C\ell_{(1-n)}$ is the double cover $\rm{Pin}(n-1)$ of the full orthogonal group $\rm{O}(n-1)$ and its irreducible representations are called "pinors" - see \cite{B}, (Sect.2.3).

In summary, the alternative algebras are classified as follows (\cite{R08} Proposition 1.6):

\begin{theorem} Let $(\mathcal{A}, N)$ be a composition algebra. For $\mu= \pm 1$, denote  by $\mathcal {A}(\mu)$ the algebra $\mathcal {A}(\mu)=\mathcal{A} \oplus e\mathcal{A}$ with $e^2=-\mu$  and product (\ref{1.16}). Then
\begin{itemize}
\item $\mathcal{A}(\mu)$ is commutative iff $\mathcal {A}=\mathbb{K}$;
\item $\mathcal {A}(\mu)$ is associative iff $\mathcal {A}$ is associative and  commutative;
\item $\mathcal{A}(\mu)$ is alternative iff $\mathcal{A}$ is  associative.
\end{itemize}
\end{theorem}
\begin{theorem} (\cite{R08} Theorems (\ref{1.7})-(\ref{1.10})). A composition algebra is, as a vector space, $1,2,4$ or $8$  dimensional. There are four composition algebras $\mathcal {A}_j$ over $\mathbb{C}$ of dimension $2^j$, $j=0,1,2,3$. There are seven composition algebras over $\mathbb{R}$; the division algebras $\mathcal {A}_j^+$, $(j=0,1,2,3)$ with $N(x)\geq 0 $ and
$x^{-1}=\frac{x^*}{N(x)}$ for $x \neq 0$, and the split algebras $\mathcal {A}_j^s$, $j=1,2,3$ of signature $(2^{j-1},2^{j-1})$.
\end{theorem}

All above algebras are unique up to isomorphism. The multiplication rule (\ref{1.16}) varies in different expositions. Different conventions are related by algebra automorphisms. (Our notation differs from Roos \cite{R08} only by the sign of $\mu$, as we set $e^2=-\mu$.) The only nontrivial automorphism of the algebra of complex numbers is the complex conjugation. The automorphism group of the (real) quaternions is $SO(3)$ realized by
\begin{eqnarray}
x \rightarrow uxu^* ,\qquad u \in SU(2) \ (u^*=u^{-1}).
\label{1.25}
\end{eqnarray}
Similarly, the automorphism group of the split quaternions is $SO(2,1)$:
 \begin{eqnarray}
 \mathbb{H} \ni x\rightarrow gxg^{-1}, \qquad g \in SL(2,\mathbb{R}).
 \label{1.26}
\end{eqnarray}
We shall survey the octonions and their automorphisms in the next section.

\subsection{Historical note}
The simplest relation of type (\ref{1.1}), the one applicable to the absolute value square of a product of complex numbers
$$(xu-yv)^2+(xv+yu)^2=(x^2+y^2)(u^2+v^2)$$
$(x,y,u,v \in \mathbb{R})$, was found by Diophantus of Alexandria around 250. A more general relation of this type,
$$(xu+Dyv)^2-D(xv+yu)^2=(x^2-Dy^2)(u^2-Dv^2)$$
occurs for special values of $D$ in Indian mathematics (Brahmegupta 598) - see \cite{B61} Sect.2. For $D$ positive it applies to the \textit{split complex numbers}. The geometric interpretation by Gauss comes much later. (The fact that complex numbers are useful and should be taken seriously is sometimes attributed to Gerolamo Cardano (1501-1576), whose book Ars Magna (The Great Art) contains the solution of the cubic equation. In fact, it was his contemporary, Bologna's mathematician Rafael Bombelli (1526-1572) who first thoroughly understood the complex numbers and described them in his L'Algebra, published in 1572.)

The multiplicativity of the norm of the quaternions was noted by Euler in 1748, a century before Hamilton discovered the algebra of quaternions in 1843 (when "in a famous act of a mathematical vandalism, he carved the equations $i^2=j^2=k^2=ijk=-1$ into the stone of Brougham Bridge" \cite{B} p. 145). The corresponding relation for the octonions was discovered by the Danish mathematician Degen in 1818 - again before the discovery of the octonions (which took place in late 1843 - in a letter to Hamilton by his college friend J.T.Graves). The first publication about octonions appears as an appendix to an otherwise erroneous paper of the English mathematician (at the time, lawyer) Arthur Cayley (1821-1895) in 1845 (see Introduction and references  $17,18$ of \cite {B})

The American algebraist and author of a three-volume \textit{History of the Theory of Numbers}, Leonard E. Dickson (1874-1954) contributed to the construction of composition algebras in 1919 \cite{Di}. The theorem that the only normed division algebras are $\mathbb{R}$, $\mathbb{C}$, $\mathbb{H}$  and   $\mathbb{O}$ was proven by A. Hurwitz (1859 -1919) in 1898. The extension of this result to alternative  (including split) algebras belongs to M. Zorn (1906-1993) in 1930 and 1933. The fact that the only division algebras (without extra structure) have dimensions $1,2,4,8$ was established as late as in 1958 (independently by R. Bott and J. Milnor and by M. Kervaire).

\bigskip

\section{Octonions. Isometries and automorphisms}
\setcounter{equation}{0}
\renewcommand\theequation{\thesection.\arabic{equation}}
\subsection{Eight dimensional alternative algebras}
The multiplication table of the imaginary octonions (see Appendix A) can be introduced by first selecting a quaternion subalgebra
\begin{eqnarray}
e_je_k= \epsilon_{jkl}e_l-\delta_{jk},\qquad j,k,l =1,2,4,\nonumber \\ \epsilon_{124}=1=\epsilon_{412}=\epsilon_{241}=1=-\epsilon_{214}=\dots .
\label{2.1}
\end{eqnarray}
The somewhat exotic labeling of the units (jumping over $3$) is justified by the following memorable multiplication rules $mod {7}$:
\begin{eqnarray}
e_ie_j=e_k \Rightarrow e_{i+1} e_{j+1}=e_{k+1}, \qquad e_{2i}e_{2j}=e_{2k}\equiv e_{2k(mod {7})} \nonumber \\
e_7e_j=e_{3j(mod {7})}, \quad \text{for} \ j=1,2,4 \  (e_7e_4=e_5).
\label{2.2}
\end{eqnarray}
A convenient complex isotropic basis for the vector representation of the isometry Lie algebra $so(8)$ of $\mathbb{O}$ (which contains the authomorphism algebra $\mathfrak{g}_2$ of the octonions) is given by:
\begin{eqnarray}
\rho^{\epsilon}=\frac{1}{2}(1 +i\epsilon e_7), \quad
\zeta_j^{\epsilon}=\rho^\epsilon e_j = \frac{1}{2}(e_j+i\epsilon e_{3j}) \quad  j=1,2,4, \ \epsilon=\pm
\label{2.3}
\end{eqnarray}
(the imaginary unit $i$ commutes with the octonion units $e_a$).
The multiplication table of the octonion units is summarized by the following relations:
\begin{eqnarray}
(\zeta_j^{\epsilon})^2&=&0=\rho^{+}\rho^{-}, \quad (\rho^{\epsilon})^2=\rho^{\epsilon}, \ \rho^{+}+\rho^{-}=1,  \quad
\zeta_j^{\epsilon}\zeta_k^{\epsilon}=\epsilon_{jkl}\zeta_l^{-\epsilon}\nonumber \\ \zeta_j^{\epsilon}\zeta_k^{-\epsilon}&=&-\rho^{-\epsilon}\delta_{jk},\quad \Rightarrow [\zeta_j^+, \zeta_k^-]_+ =\delta_{jk}, \quad  j,k,l=1,2,4.
\label{2.4}
\end{eqnarray}
The idempotents $\rho^\pm$ (which go back to G\"ursey) are also exploited in \cite{D10}. The last equation (\ref{2.4}) coincides
with the canonical anticommutation relations for fermionic creation and annihilation operators (cf. \cite{CGKK}).

The \textit{split octonions} $x_s\in\mathbb{O}_s$ with units $\tilde{e}_a$ can be embedded in the algebra $\mathbb{C}\mathbb{O}$ of complexified octonions by
setting $\tilde{e}_{\mu}=e_{\mu}$, $\mu=0,1,2,4$; $\tilde{e}_7=ie_7$,
 $\tilde{e}_{3j}=ie_{3j(mod 7)}$,  so that
\begin{eqnarray}
x_s&=&\sum_{a=0}^7x_s^a\tilde{e}_a \Rightarrow N(x_s)=x_sx_s^*  \nonumber \\
&=&\sum_{\mu=0,1,2,4}(x_s^{\mu})^2-(x_s^7)^2-(x_s^3)^2-(x_s^6)^2-(x_s^5)^2.
\label{2.5}
\end{eqnarray}
The quark-lepton correspondence suggests the splitting of octonions into a direct sum,
\begin{eqnarray}
\mathbb{O}=\mathbb{C} \oplus \mathbb{C}^3; \ x = a +\textbf{z e} =a+z^1e_1+z^2e_2+z^4e_4  \  (e_1e_2=e_4)  \nonumber \\
a=x^0+x^7e_7, \  z^j=x^j+x^{3j(mod7)}e_7  \ (x^{12}\equiv x^5)
\label{2.6}
\end{eqnarray}
thus $e_7$ playing the role of \textit{imaginary unit within the real octonions}. The Cayley-Dickson construction corresponds to the splitting of $\mathbb{O}$ into two quaternions:
\begin{eqnarray}
 \mathbb{O}=\mathbb{H} \oplus\mathbb{H}:
x=u+e_7v, \ u=x^0+x^je_j, \  v=x^7+x^{3j}e_j.
\label{2.7}
\end{eqnarray}
One may speculate that upon complexification $u$ and $v$ could be identified with the "up-" and "down-", leptons and quarks: $u=(\nu; u^j, \ j=1,2,4)$. $v=(e^-, d_j)$ $j$ playing the role of a colour index, but we shall not pursue this line of thought. For $x_r=u_r+e_7v_r$, \ $r=1,2$ the Cayley-Dickson formula (\ref{1.16}) and its expression in terms of the complex variable $a_r, z_r^j$ reads:
\begin{eqnarray}
x_1x_2=u_1u_2-v_2v_1^* +e_7(u_1^*v_2+u_2v_1)=  \nonumber  \\
a_1a_2 -\mathbf{z}_1\mathbf{\overline{z}_2}+(a_1\mathbf{z_2}+\overline{a}_2\mathbf{z}_1+\mathbf{\overline{z}_1}\times \mathbf{\overline{z}_2})\mathbf{e}
\label{2.8}
\end{eqnarray}
where the star indicates quaternionic conjugation while the bar stands for a change of the sign of $e_7$ $(\bar{z}^j =x^j-e_7x^{3j})$. The two representations (\ref{2.8}) display the covariance of the product under the action of two subgroups of maximal rank of the automorphism group of the octonions. If $p$ and $q$ are two unit quaternions
\begin{eqnarray}
p&=& p^0+p^je_j, \qquad  q=q^0+q^je_j \nonumber \\  pp^*&=&N(p)=(p^0)^2+\mathbf{p^2}
=1=qq^*  \Leftrightarrow  (p,q)\in SU(2)\times SU(2),
\label{2.9}
\end{eqnarray}
it is easy to verify, using the first equation (\ref{2.8}), that the transformation
\begin{eqnarray}
(p,q): x=u+e_7v,\ \rightarrow pup^* +e_7pvq^*,  \nonumber \\
(p,q) \in \frac{SU(2) \times SU(2)}{\mathbb{Z}_2^{diag}}
\label{2.10}
\end{eqnarray}
where $\mathbb{Z}_2^{diag}=\{(p,q)=(1,1), (-1,-1)\}$, is an automorphism of the octonion algebra. Similarly, if $U \in SU(3)$ acts on $x$  (\ref{2.6}) as
\begin{eqnarray}
U: x=a+z^je_j \rightarrow a+U^i_jz^je_i \  \text{then} \ U(x_1)U(x_2) = U(x_1x_2).
\label{2.11}
\end{eqnarray}

The subgroups (\ref{2.10}), (\ref{2.11}) are the two closed connected subgroups of maximal rank of the compact group $G_2$ corresponding to the \textit{Borel - de Siebenthal theory} \cite{BdS} that plays a central role in \cite{TD} as well as in Chapter 4 below.

\subsection{Isometry group of the (split) octonions. Triality}
The norm $N(x)$ (\ref{1.17}) and the associated scalar product of the (split) octonions,

\begin{eqnarray}
N_{(s)}(x)=\sum_{a=0}^7\eta_a^{(s)}x_a^2, \quad \eta_a=1 \quad \text{for all}\quad a,\quad \text{for}\quad x\in \mathbb{O} \nonumber \\
\eta_0^s=\eta_1^s=\eta_2^s=\eta_4^s=1=-\eta_7^s=-\eta_3^s=-\eta_6^s=-\eta_5^s,\quad \text{for} \quad x\in\mathbb{O}_s
\label{2.12}
\end{eqnarray}
is the (compact) orthogonal group $\rm{O}(8)$ (respectively, the split orthogonal group $\rm{O}(4,4)$). As stressed in \cite{B}, while invariant quadratic forms are common, trilinear forms are rare. It is, therefore, noteworthy, that the trilinear form $t(xyz)$ (\ref{1.10}) is invariant under the $2$-fold cover $\rm{Spin}(8)$ (respectively $\rm{Spin}(4,4)$) of the connected orthogonal group $\rm{SO}(8)$ (respectively, $\rm{SO}(4,4)$).
The existence of such trilinear invariant is related to the exceptional symmetry of the Lie algebra $\frak{D}_4=\frak{so}(8)$ which is visualized by the symmetry of its Dynkin diagram, Figure 1.
\newpage
\begin{figure}
	\centering
	\includegraphics[scale=0.7]{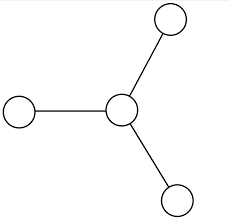}
	\caption{The Dynkin diagram of the Lie algebra $\frak{D}_4$. The three outer nodes correspond to the two $8$-dimensional chiral spinor representations $S_8^{\pm}$ of $\rm{Spin}(8)$ and to the vector representation of $\rm{SO}(8)$, the central node corresponds to the $28$-dimensional adjoint representation. The \textit{triality} automorphisms implementing the symmetries of the diagram were discovered in 1925 by \'Elie Cartan \cite{Ca}.}
	\label{fig:The-Dynkin-diagram-of-SO8}
	%\end{centering}
\end{figure}
Accordingly, the symmetric group $\rm{S}_3$ which permutes the three external nodes of  $\frak{D}_4$ is the group of outer automorphisms of the Lie algebra $\frak{so}(8)$.
This is the \textit{triality group} of $\rm{D}_4$.
We proceed to describing the Lie algebra $\frak{D}_4$  and to displaying the action of the outer automorphisms on it.
To begin with $\frak{D}_4$ is generated by left multiplications $L_{\alpha}$ (as well as by right multiplications $R_{\alpha}$) by pure imaginary elements $\alpha  \in \rm{Im} \mathbb{O}$ acting on the algebra of octonions. If we set (following \cite{M89}, \cite{Y})
\begin{eqnarray}
L_{\alpha}x=\alpha x, \qquad R_{\alpha} x=x\alpha, \qquad T_{\alpha} = L_{\alpha}+R_{\alpha} \nonumber \\
(\text{i.e.} \quad T_{\alpha}x=\alpha x+x\alpha) \qquad \text{for} \qquad \rm{Re}\alpha=<\alpha,1>=0)
\label{2.13}
\end{eqnarray}
then each of the three actions annihillates the inner product:
\begin{eqnarray}
<L_{\alpha}x,y>+<x,L_{\alpha}y>=0 \quad \text{etc. since}\quad \alpha^*=-\alpha.
\nonumber
\end{eqnarray}
Note that the product of operators $L_a L_b$, $a,b\in \mathbb{O}$ cannot be written in general as $L_{ab}$ (or as left multiplication by any element of $\mathbb{O}$). But, as emphasized in the thesis \cite{F16} one can work with composition of maps $L_aL_b$ which is associative while the product of octonions is not. In fact, the Lie algebra $\frak{so}(8)$ is spanned by $L_{e_i}$ and  $[L_{e_j},L_{e_k}]$  for $1\leq i,j, k\leq 7$. The action of the group $\rm{S}_3$ of outer automorphism
of $\frak{D}_4$ is generated by an automorphism of order three $\nu$ and an involution $\pi$ defined on the triple $(L_{\alpha},R_{\alpha},T_{\alpha})$ by
\begin{eqnarray}
\nu L_{\alpha}&=& R_{\alpha}, \qquad \nu R_{\alpha}(=\nu^2 L_{\alpha})=-T_{\alpha}\quad \Rightarrow \quad \nu T_{\alpha}=-L_{\alpha}, \qquad \nu^3=id \nonumber \\
\pi L_{\alpha}&=&T_{\alpha}, \qquad \pi R_{\alpha}=-R_{\alpha}\quad \Rightarrow \quad \pi T_{\alpha}=L_{\alpha}, \qquad \pi^2=id.
\label{2.14}
\end{eqnarray}
We leave it as an exercise to the reader to verify that the product $\kappa =\pi \nu$ is an involution ($\kappa^2=id$) satisfying
\begin{eqnarray}
\kappa L_{\alpha}= -R_{\alpha}, \quad \kappa R_{\alpha} =-L_{\alpha},  \quad \kappa T_{\alpha}=-T_{\alpha}, \quad \nu \kappa=\pi, \quad \nu \pi=\pi \nu^2.
\label{2.15}
\end{eqnarray}
The involutive automorphism $\kappa$ can be extended to an arbitrary element $D$ of the Lie algebra $\frak{D}_4$ with the formula (\cite{Y} Sect.1.3).
\begin{eqnarray}
(\kappa D)x=(Dx^*)^*,\quad \text{for all}\quad D\in \frak{D}_4=\frak{so}(8), \quad x\in \mathbb{O}.
\label{2.16}
\end{eqnarray}
We shall introduce,  following\footnote{We note that the convention $e_1e_2=e_3$ (rather than our $e_1e_2=e_4$ which followss \cite{B, CS}) is used in \cite{M89, Y} so that our formulas relating $G$ and $F$ (Appendix B) differ from theirs.} \cite{M89, Y}, two bases, $G_{ab}$ and $F_{ab}$, in $\frak{D}_4$, defined by
\begin{eqnarray}
G_{ab} e_c&=&\delta_{bc} e_a-\delta_{ac} e_b,\qquad a,b,c=0,1,\dots,7, \qquad e_0=1 \nonumber \\
F_{ab}x&=&\frac{1}{2}e_a(e_b^*x)=-F_{ba}x \qquad (F_{aa}:=0, \quad e_0^*=e_0)
\label{2.17}
\end{eqnarray}
which satisfy the same commutation relations:
\begin{eqnarray}
\left[X_{ab},X_{cd}\right]=\delta_{bc}X_{ad}-\delta_{bd}X_{ac}+\delta_{ad}X_{bc}-\delta_{ac}X_{bd}, \quad X=G,F.
\label{2.18}
\end{eqnarray}
Using the identity $L_{e_k}=2F_{k0}$ (for $k=1,\dots,7$) and Eq. (\ref{2.14}) one verifies that $\pi(F_{k0})=G_{k0}$. More generally, we have
\begin{eqnarray}
\pi(G_{ab})=F_{ab}, \qquad \pi(F_{ab})=G_{ab}
\label{2.19}
\end{eqnarray}
(cf. Lemma 1.3.3 of \cite{Y}).
We shall demonstrate in Appendix B that the involution $\pi$ splits into
seven $4$-dimensional involutive transformations. Here, we display one of them which involves our choice of the Cartan subalgebra of $so(8)$:
\begin{eqnarray}
F_7&=&X_7G_7,  \,\ G_7= \left( \begin{array}{c}
G_{07} \\
G_{13} \\
G_{26} \\
G_{45} \\
\end{array} \right),
F_7= \left( \begin{array}{c}
F_{07} \\
F_{13} \\
F_{26} \\
F_{45} \\
\end{array} \right), \nonumber  \\
X_7&=&\frac{1}{2} \left( \begin{array}{cccc}
1 &  1 &  1 &  1 \\
1 &  1 & -1 & -1 \\
1 & -1 &  1 & -1 \\
1 & -1 & -1 &  1 \\
\end{array} \right),
\label{2.20}
\end{eqnarray}
a straightforward calculation gives $X_7^2=\un$, $\det X_7=-1$.

\begin{theorem}
(Infinitesimal triality) For any $D\in \frak{D}_4$ and $x,y \in \mathbb{O}$ the following identity holds:
\end{theorem}
\begin{eqnarray}
(Dx)y+x\nu(D)y=\pi(D)(xy).
%\eqno \rm{(2.21)}$$
\label{2.21}
\end{eqnarray}
%\end{theorem}
\textit{For a given} $D\neq0$ \textit {the automorpisms $\nu$ and  $\pi$}, \textit{given by} (\ref{2.14}), (\ref{2.19}) \textit{are uniquely determined from}  \rm{(2.21)}.	

For $D=L_{\alpha}$ ($\rm{Re}\alpha=0$) Theorem 3 follows from the definition (\ref{2.14}). For a general $D\in \rm{D}_4$ one uses the fact that it can be written as a linear combination of $L_{\alpha}$ and their commutators (see Theorem 1.3.6 of \cite{Y}).

There exists a group theoretic counterpart of the principle of infinitesimal triality which we shall only sketch. (For a pedagogical exposition based on the concepts of Moufang loops and their isotopies - see Chapters 7 and 8 of \cite{CS}.)

The notion of left and right multiplication are well defined and preserve the norm for unit octonions; for instance $N(L_{a}x)= N(x)$ for $aa^*=1$. In order to obtain a general $\rm{SO}(8)$ action on the octonions one needs seven left (or seven right) multiplications (\cite{CS} Sect. 8.4). The group theoretic counterpart of $T_{\alpha}$ is the bimultiplicaton $B_{a}x=axa$; for $a_t=\exp(t\alpha)$,  $t\in \mathbb{R}$, $\alpha \in \rm{Im}\mathbb{O}$
\begin{eqnarray}
\lim_{t\rightarrow 0} \frac{d}{dt}B_{a_t}x=T_{\alpha}x=\alpha x+x\alpha. %\rm{(2.22)}$$.
\label{2.22}
\end{eqnarray}
The automorphusms (\ref{2.14}), (\ref{2.15}) can be lifted to group automorphisms:
\begin{eqnarray}
\nu(L_a,R_a,B_a)&=&(R_a,B_{a^*},L_{a^*}),\qquad \pi(L_a,R_a,B_a)=(B_a,R_{a^*},L_a) \nonumber \\
\kappa(L_a,R_a,B_a)&=&(R_{a^*},B_{a^*},L_{a^*}), \qquad \kappa^2=\pi^2=\nu^3=1.
\label{2.23}
\end{eqnarray}
\begin{theorem}
 For each rotation $g\in \rm{SO}(8)$ there are ellements $g^{\pm}$ of the double cover $\rm{Spin}(8)$ of $\rm{SO}(8)$ such that the normed trilinear form
 \end{theorem}
 \begin{eqnarray}
 t_8(x,y,z)=\frac{1}{2}t(x,y,z)=<xyz,1>, \qquad x,y,z\in \mathbb{O}
 %\eqno \rm{(2.24)}$$
 \label{2.24}
 \end{eqnarray}
 \textit{satisfy the invariance condition}
  \begin{eqnarray}
 t_8(gx,g^+y,g^-z)=t_8(x,y,z).
 \label{2.25}
 \end{eqnarray}
 %\eqno \rm{(2.25)}$$
 \textit{The elements} $g^{\pm}$ \textit{are determined from} (\ref{2.25}) \textit{up to a common sign: for each} $g$ \textit{there are exactly two pairs} $(g^+,g^-)$ \textit{and} $(-g^+,-g^-)$ \textit{obeying} (\ref{2.25}). \textit{For} $g=L_a \quad (aa^*=1)$ \textit{we can set} ,
 \begin{eqnarray}
 g^+=\nu(L_a)=R_a, \qquad g^-=\nu^2L_a=B_{a^*}.
 %\eqno \rm{(2.26)}$$
\label{2.26}
\end{eqnarray}
 (The factor $\frac{1}{2}$ in (\ref{2.24}) follows the definitionn of \textit{normed triality}   of \cite{B} which demands $|t_8(x,y,z|^2 \leq N(x)N(y)N(z).)$
\subsection{Automorphism group $\rm{G}_2$ of the octonions.}
% and its maximal subgroup}
As proven by \'Elie Cartan in 1914, the automorphism group of the octonions is the rank $2$ exceptional Lie group $\rm{G}_2$ which can thus be defined by
\begin{eqnarray}
G_2=\{g\in L(\mathbb{O},\mathbb{R})|(gx)(gy)=g(xy), \ x,y \in \mathbb{O}\}
\label{2.27}
\end{eqnarray}
where $ L(\mathbb{O},\mathbb{R})$ is the group of non-singular linear transformations of the $8$-dimensional real vector space
$\mathbb{O}$. It follows from $e_0^2=e_0=1$ and $e_k^*=-e_k$ for $k>0$ that $\rm{G}_2$ preserves the octonion unit and the Cayley conjugation and hence the   norm $N(x)$:
\begin{eqnarray}
g1 =1, \qquad (gx)^* =g(x^*), \qquad N(gx)= N(x).
\label{2.28}
\end{eqnarray}
Thus $\rm{G}_2$ is a subgroup of the isometry (orthogonal) group $\rm{O}(\mathbb{O})=\rm{O}(8)$  of the $8$-dimensional euclidean space of the octonions. In fact, it is a subgroup of the connected orthogonal group $\rm{SO}(7)$ of the $7$-dimensional space $\rm{Im}\mathbb{O}$ of imaginary octonions, the Lie algebra $\frak{so}(7)$ splitting as a vector space into a direct sum of the Lie algebra $\mathfrak g_2$ and the vector representation \underline{7} of $\frak{so}(7)$. Thus the dimension of $\rm{G}_2$ is $\binom{7}{2}-7=14$.
%and its lowest order faithful $7$-dimensional representation:
\begin{eqnarray}
so(7)\approxeq \mathfrak  g_2 \oplus \underline{7} \,\ (=\mathfrak  g_2 \oplus\mathbb{R}^7).\nonumber
%\label{2.29}
\end{eqnarray}
%We see, in particular, that the dimension of $G_2$ is $14$.
Moreover, the group $\rm{G}_2$ acts transitively on the unit sphere $\mathbb{S}^6$ in $\mathbb{R}^7$; every point of $\mathbb{S}^6$ can be transformed, say, into $e_7$ by an authomorphism of $\mathbb{O}$, The stabilizer of $e_7$ is the subgroup  $\rm{SU}(3)$ of $\rm{G}_2$, defined by (\ref{2.11}):
\begin{eqnarray}
(\rm{G}_2)_{e_7}= \rm{SU}(3) \Rightarrow \frac{\rm{G}_2}{\rm{SU}(3)} \approxeq \mathbb{S}^6.
\label{2.29}
\end{eqnarray}
It follows that the group $\rm{G}_2$ is connected.
The maximal subgroups of $G_2$, whose action was defined by (\ref{2.10}) and (\ref{2.11}) (and which correspond to the Borel - de Siebenthal theory) can be characterized as follows. The \textit{complex conjugation} $\gamma$ (in the notation of \cite{Y}): $e_7 \rightarrow -e_7$ belongs to the automorphism group $G_2$ of $\mathbb{O}$ (corresponding, in fact, to the reflection of four imaginary units $e_7, \ e_7e_1=e_3, \ e_7e_2=e_6 \,\ \text{and} \,\ e_7e_4=e_5$) and has square one:
\begin{eqnarray}
\gamma x =\gamma(u+e_7v)=u-e_7v, \ \gamma^2=1.
\label{2.30}
\end{eqnarray}
The rank two  (semisimple) subgroup (\ref{2.9}), (\ref{2.10}) of $G_2$ can be characterized as the commutant of $\gamma$ in $G_2$:
\begin{eqnarray}
G_2^{\gamma}=\{g \in G_2| \gamma g=g \gamma \} = \frac{SU(2)\times SU(2)}{\mathbb{Z}_2}.
\label{2.31}
\end{eqnarray}
Denote, on the other hand, by $\omega$ the generator of the center of $SU(3)$  acting on $z$ by (\ref{2.11})
\begin{eqnarray}
\omega x=a +\omega_7 z^je_j, \,\ \omega_7=-\frac{1}{2}+\frac{\sqrt 3}{2}e_7 \,\ (\omega_7^3=1=\omega^3).
\label{2.32}
\end{eqnarray}
Then the subgroup  (\ref{2.11}) of $G_2$ is characterized by
\begin{eqnarray}
G_2^{\omega}:=\{\omega \in G_2|\omega g=g\omega\}=SU(3)\left(=(\rm{G}_2)_{e_7}\right).
\label{2.33}
\end{eqnarray}
(One uses, in particular, the relation $\omega_7\mathbf{z}\mathbf{e}\omega_7=\omega_7\overline{\omega}_7\mathbf{z}\mathbf{e}=\mathbf{z}\mathbf{e}$.)

\subsection{Roots and weights of ${\mathfrak g}_{2}\subset\frak{B}_3
	\subset\frak{D}_4 $}
It is convenient (in particular, for the study of the Jordan algebra $JSpin_9$ in Sect. 3.3 below to view the Lie algebra $\frak{D}_4$ as embedded into the Clifford algebra $C\ell_8$). Indeed, its $16\times16$ matrix-generators $\hat{P}$, $a=0,1,\dots,7$, satisfy the same anticommutation relations as the $2\times2$ hermitian octonionic matrices
\begin{eqnarray}
\hat{e}_a= \left(
\begin{array}{cc}
0 &  e_a     \\
e_a^*    & 0\\
\end{array}\right), \ \text{i.e.} \
\hat{e}_0= \left(
\begin{array}{cc}
0 &  e_0     \\
e_0    & 0\\
\end{array}\right),  (e_0=1),\
\hat{e}_k= \left(
\begin{array}{cc}
0 &  e_k     \\
-e_k    & 0\\
\end{array}\right),
\label{2.34}
\end{eqnarray}
 $k=1,\dots,7$; we have
\begin{eqnarray}
[\hat{e}_a,\hat{e}_b]_+=2\delta_{ab}\mathfrak{\un}_2 \qquad\leftrightarrow \qquad \left[\hat{P}_a,\hat{P}_b\right]_+=2\delta_{ab}\mathfrak{\un}_{16},\quad a, b=0,1,\dots,7 .
\label{2.35}
\end{eqnarray}
(The symbol $1_n$, the $n\times n$ unit matrix, will henceforth be omitted wherever this  would not give rise to ambiguity.) The Clifford algebra gives also room for the symmetry generators. If we set
\begin{eqnarray}
\hat{G}_{ab}= \left(
\begin{array}{cc}
G_{ab} &  0     \\
0    & G_{ab}^{\kappa}
\end{array}\right),\qquad \text{where}\qquad  G_{ab}^{\kappa}=\kappa(G_{ab}),
\label{2.36}
\end{eqnarray}
so that, in view of (\ref{2.16}) and (\ref{2.17}),
$G^{\kappa}_{ab}\hat{e}_c^*=\delta_{bc}e_a^*-\delta_{ac}e_b^*$
we shall have
\begin {eqnarray}
\hat{G}_{ab}\hat{e}_c=\delta_{bc}\hat{e}_a-\delta_{ac}\hat{e}_b, \qquad a,b,c=0,1,\dots,7.
\label{2.37}
\end{eqnarray}
Writing the $\hat{P}$, in analogy with (\ref{2.34}), as
\begin{eqnarray}
\hat{P}_0&=& \sigma_1\otimes P_0= \left(
\begin{array}{cc}
0   &  P_0     \\
P_0 &  0       \\
\end{array}\right), \quad
\hat{P}= c\otimes P_k=\left(
\begin{array}{cc}
 0    &  P_k   \\
-P_k  &  0     \\
\end{array}\right), \   k=1,\dots,7,  \nonumber \\
 P_0&=&\un_8,\ c=i\sigma_2, \quad \left[P_j,P_k\right]_+=-2\delta_{jk},\ P_1P_2\dots P_6=P_7,
\label{2.38}
\end{eqnarray}
we see that the generators $G_{ab}$ of the Lie algebra $\frak{D}_4$ are represented by
\begin{eqnarray}
\frac{1}{2}\hat {P}_{ab}=\frac{1}{4}\left[\hat{P}_a,\hat{P}_b\right],\ \left[\frac{1}{2}\hat{P}_{ab}, \hat{P}_c\right]=\delta_{bc}\hat{P}_a-\delta_{ac}\hat{P}_b, \ a,b,c=0,1,\dots,7.
\label{2.39}
\end{eqnarray}
We stress that the map $\hat {e}_a\rightarrow \hat{P}_a$ (and hence the map $\hat{e}_a\rightarrow \Gamma_a$) - unlike the representation $q_j\rightarrow -i\sigma_j$ of the imaginary quaternion units - only respects the anticommutators, the relation $L_{e_1}...L_{e_6}1=e_7$, and the $\rm{SO}(\mathbb{O})$ symmetry properties of the octonions, not their commutators.

The four operators ${G}_7$ of (\ref{2.20}) are diagonalized in the counterpart of the isotropic octonionic basis (\ref{2.3}), (\ref{2.4})
given by
\begin{eqnarray}
\hat{P}^{\epsilon}_0&=&\frac{1}{2}(\hat{P}_0 +i\epsilon \hat{P}_7), \quad
\hat{P}^{\epsilon}_j=\frac{1}{2}(\hat{P}{j} + i\epsilon \hat{P}_{3j}),\ j=1,2,4, \ (3\times4=5(mod7))   \nonumber \\
\epsilon & =&\pm, \ \text{s.t.} \ 
(\hat{P}^{\epsilon}_{\mu})^2=0, \ 
\hat{P}^{-\epsilon}_{\mu}\hat{P}^{\epsilon}_{\mu}=:\rho_{\mu}^{\epsilon}, \quad (\rho_{\mu}^{\epsilon})^2 =\rho _{\mu}^{\epsilon}, \ \rho_{\mu}^+ +\rho_{\mu}^-=\un_{16}.
\label{2.40}
\end{eqnarray}
Their commutators define an orthogonal weight basis in $\frak{D}_4 $ (as well as in $\frak{B}_4 =\frak{so}(9)$ - see Sect. 4.1):
\begin{eqnarray}
\Lambda_{\mu}=\frac{1}{2}\left[\hat{P}_{\mu}^-,\hat{P}_{\mu}^+\right]=\frac{1}{2}(\rho_{\mu}^+ -\rho_{\mu}^-),\ \Lambda_0=\frac{i}{2}\hat{P}_{07}, \ \Lambda_{j}=\frac{i}{2}\hat{P}_{ 3j},\ j=1,2,4,
\label{2.41}
\end{eqnarray}
which can be represented by real diagonal matrices (belonging actually to $\frak{so}(4,4)\subset C\ell(4,4))$. The operator $\Lambda_{\mu}$ are normalized to have eigenvalues $\pm 1$ (and $0$) under commutation:
\begin{eqnarray}
\left[\Lambda_{\mu},\hat{P}_{\nu}^{\epsilon}\right]=-\epsilon\delta_{\mu\nu}\hat{P}_{\nu}^{\epsilon},\quad \mu, \nu=0,1,2,4, \qquad \epsilon=\pm. %\quad\left(\rho_{\mu}^{\epsilon}\Gamma_{\mu}^{\epsilon}=0\right).
\label{2.42}
\end{eqnarray}
They also satisfy the orthogonality relation $<\Lambda_{\mu},\Lambda_{\nu}>:=\frac{1}{4}tr\left(\Lambda_{\mu},\Lambda_{\nu}\right)=\delta_{\mu\nu}$. A Chevalley-Cartan basis corresponding to the simple roots $\alpha_{\mu}$ of $\frak{D}_4$ is given by
\begin{eqnarray}
\alpha_0&\leftrightarrow &H_0=\Lambda_0-\Lambda_1 = \rho_0^+-\rho_1^+, \quad E_0=\hat{P}_0^-\hat{P}_1^+, \quad F_0=\hat{P}_1^-\hat{P}_0^+ \nonumber \\
 \alpha_1&\leftrightarrow &H_1=\Lambda_1-\Lambda_2 = \rho_1^+-\rho_2^+, \quad E_1=\hat{P}_1^-\hat{P}_2^+, \quad F_1=\hat{P}_2^-\hat{P}_1^+ \nonumber \\
\alpha_2&\leftrightarrow &H_2=\Lambda_2-\Lambda_4= \rho_2^+-\rho_4^+, \quad E_2=\hat{P}_2^-\hat{P}_4^+, \quad F_2=\hat{P}_4^-\hat{P}_2^+ \nonumber \\
(H_4=)H_{\alpha_4}&_=&\Lambda_2+\Lambda_4 =\pm \rho_2^{\pm}\mp\rho_4^{\mp}, \ E_{\alpha_4}=\hat{P}_4^-\hat{P}_2^-,  F_{\alpha_4}=\hat{P}_2^+\hat{P}_4^+
\label{2.43}
\end{eqnarray}
and satisfy the standard commutation relations
\begin{eqnarray}
\left[H_{\mu},E_{\nu}\right]=c_{\mu\nu}^DE_{\nu}, \qquad \left[H_{\mu},F_{\nu}\right]=-c_{\mu\nu}^DF_{\nu} ,\qquad [E_{\mu},F_{\nu}]=\delta_{\mu\nu}H_{\nu}
\label{2.44}
\end{eqnarray}
where $c_{\mu\nu}^D$ is the  $\frak{D}_4$ Cartan matrix
\begin{eqnarray}
(c_{\mu\nu}^D) =\left( \begin{array}{cccc}
2  & -1 &  0 & 0 \\
-1 &  2 & -1 &-1 \\
0 & -1 &  2  & 0 \\
0 &  -1 & 0  & 2 \\
\end{array} \right). \nonumber
%\label{2.18}
%\label{166}
\end{eqnarray}
As it follows from our discussion in Sect. 2.3 of the automorphism group of the octonions the derivations of $\mathbb{O}$ span, in fact, a subalgebra of the $21$-dimensional Lie algebra
\begin{eqnarray}
\frak{B}_3&=&\frak{so}_7=\{\frak{D}\in\frak{D}_4: \quad De_0=0\}=\{D\in\frak{D}_4;\quad \kappa(D)=D\} \nonumber \\
&=&\text{Span}\{G_{kl}; k,l=1,\dots,7\}.
\label{2.45}
\end{eqnarray}
The first two simple roots $\alpha_1$ and $\alpha_2$ of $\frak{B}_3$ coincide with those of $\frak{D}_4$, while the third,  $\alpha_s$, is a short root:
\begin{eqnarray}
\text{\{Simple roots of} \ \frak{B}_3\}=\{\alpha_1=\Lambda_1-\Lambda_2, \ \alpha_2=\Lambda_2-\Lambda_4, \ \alpha_s=\Lambda_4\}
\label{2.46}
\end{eqnarray}
the coroot corresponding to $\alpha_s$ being $$\alpha_s^{\vee}=\frac{2\alpha_s}{(\alpha_s,\alpha_s)}=2\Lambda_4$$
(so that   $C_{2s}^B=(\alpha_2, \alpha_s^{\vee})=-2)$.

Comparing the defining relation for a derivation $D\in \frak{g}_2$,
\begin{eqnarray}
(Dx)y+xDy=D(xy)
\label{2.47}
\end{eqnarray}
with the infinitesimal triality relation (\ref{2.21}) we conclude that $D\in\frak{g}_2$ only if $\nu(D)=\pi(D)=D$. Taking into a account the fact
that $\kappa(D)=D$ in $\frak{B}$ and that $\nu=\pi\kappa$ (according to (\ref{2.15})) we arrive at the following

\textit{Proposition} 2.3 \textit{If} $D\in\frak{B}_3$ \textit{then each of the condition} $\nu(D)=D$ \textit{and} $\pi(D)=D$ \textit{implies the other and the resulting triality invariance is necessary  and sufficient for} $D$ \textit{to belong to} $\frak{g}_2$.

Combining this result with the last equation (\ref{2.45}) we deduce that a linear combination of $G_{kl}$ belongs to $\frak{g}_2$ if it is invariant under the involution $\pi$ (\ref{2.19}). In particular, taking (\ref{2.20}) into account, we deduce that $\lambda G_{13}+\mu G_{26}+\nu G_{45} \in \frak{g}_2$ iff $\lambda+\mu+\nu=0$. In other words, the Cartan subalgebra  $\frak{g}_2$ is spanned by $G_{12}-G_{26}$ and $G_{26}-G_{45}$ whose represantatives within $\Gamma_{ab}$ are proportional to $H_1$ and $H_2$ of eq. (\ref{2.43}). More generally, using Appendix B, we find that the following seven linear combinations of $G_{lk}$ span $\frak{g}_2$
\begin{eqnarray}
	\lambda G_{24}+\mu G_{37}+\nu G_{56}, & \,\ & \lambda G_{14}-\mu G_{35}+\nu G_{76}, \nonumber \\
	\lambda G_{17}+\mu G_{25}-\nu G_{46},  &\,\  & -\lambda G_{12}+\mu G_{36}+\nu G_{75}, \nonumber  \\
	\lambda G_{16}-\mu G_{23}+\nu G_{47}, &\,\  &-\lambda G_{15}+\mu G_{27}+\nu G_{43},  \nonumber \\
	\lambda G_{13}+\mu G_{26}+\nu G_{45},  &\text{with} & \,\ \lambda +\mu + \nu =0.
	\label{2.48}
\end{eqnarray}

\smallskip

\section{Jordan algrbras and related groups}
\setcounter{equation}{0}
\renewcommand\theequation{\thesection.\arabic{equation}}

\subsection{Classification of finite dimensional Jordan algebras}
Pascual Jordan (1902-1980) the "unsung hero among the creators of quantum theory" (in the words of Schweber, 1994) asked himself in 1932 a question you would expect of an idle mathematician: Can one construct an algebra of (hermitian) observables without introducing an auxiliary associative product? He arrived, after some experimenting with the \textit{special Jordan product}
\begin{eqnarray}
A \circ B=\frac{1}{2}(AB+BA) \ (=B\circ A),
\label{3.1}
\end{eqnarray}
at two axioms (Jordan, 1933)
\begin{equation}
(i): \  A \circ B=B \circ A; \,\ (ii): \ A^2 \circ (B\circ A)=(A^2\circ B)\circ A,
%\label{J1-2}
\label{3.2}
\end{equation}
where $A^2:=(A\circ A)$. They imply, in particular, power associativity and
\begin{eqnarray}
A^m\circ A^n=A^{m+n}, \ m,n=0,1,2, ... .
\label{3.3}
\end{eqnarray}
(Jordan algebras are assumed to contain a unit and $A^0=1$.)  Being interested in extracting the properties of the algebra of hermitian matrices (or selfadjoint operators) for which $A^2\geq 0$, Jordan adopted Artin's \textit{formal reality} condition
\begin{eqnarray}
A_1^2+ \dots +A_n^2=0 \Longrightarrow A_1=0=\dots =A_n.
\label{3.4}
\end{eqnarray}
(It is enough to demand (3.4) for n=2.) Algebras over the real numbers satisfying both (\ref{3.2}) and (\ref{3.4}) are now called \textit{euclidean Jordan algebras}. In a fundamental paper of 1934 Jordan, von Neumann and Wigner \cite{JvNW} classified all finite dimensional euclidean Jordan algebras. They split into a direct sum of \textit{simple algebras}, which belong to four infinite families,
\begin{eqnarray}
 \mathcal{H}_n(\mathbb{R}), \quad  \mathcal{H}_n(\mathbb{C}), \quad \mathcal{H}_n(\mathbb{H}),\quad JSpin_n,
\label{3.5}
\end{eqnarray}
 and a single exceptional one
\begin{eqnarray}
\mathcal{J} (=J^8_3)=\mathcal{H}_3(\mathbb{O}).
\label{3.6}
\end{eqnarray}
Here $\mathcal{H}_n(\mathbb{A})$ stands for the set of $n\times n$ hermitian matrices with entries in the division ring $ \mathbb{A}(=\mathbb{R},\mathbb{C},\mathbb{H},\mathbb{O})$, equipped with the commutative product (\ref{3.1}). (One uses the same notation when $\mathbb{A}$ is replaced by one of the alternative split composition rings, $\mathbb{C}_s,\mathbb{H}_s$ or $ \mathbb{O}_s$ albeit the resulting algebra is not euclidean in that case.) $JSpin_n$  is an algebra of elements ($\xi,x;\xi \in \mathbb{R}, x\in\mathbb{R}^n$) where $\mathbb{R}^n$ is equipped with the (real) euclidean scalar product $<x,y>$ and the product in $JSpin_n$  is given by
\begin{eqnarray}
(\xi,x)(\eta,y)=(\xi \eta +<x,y>,  \xi y+\eta x).
\label{3.7}
\end{eqnarray}

The first three algebras $\mathcal{H}_n(\mathbb{A})$ (\ref{3.5}) are \textit{special}: the matrix product $AB$ in (\ref{3.1}) is  \textit{associative}. The algebra $JSpin_n$  is also special as a Jordan subalgebra of the $2^n$ dimensional (associative) Clifford algebra $C\ell_n$.

\textit{Remark} 3.1. The Jordan algebras $\mathcal{H}_2(\mathbb{A})$ for
$\mathbb{A}= \mathbb{R}, \  \mathbb{C}, \ \mathbb{H}, \ \mathbb{O}$ are isomorphic to  $JSpin_n$ for $n=2,3,5,9$, respectively. In fact, more generally, let $x=x^0+\mathbf{x}$ where $\mathbf{x}=\sum_{k=1}^{n-2} x^ke_k
\in C\ell_{2-n}$; then the two-by-two (Clifford valued) matrix:
\begin{eqnarray}
X&=&\left(\begin{array}{cc}
\xi+x_{n-1}  & x  \\
 x^*     & \xi-x_{n-1}    \\
\end{array} \right)
\label{3.8}
\end{eqnarray}
satisfies Eq. (\ref{1.5}) with $N(X)=\det X, \quad t(X)=\rm{tr}(X)=2\xi$ where the determinant of $X$ has a (time-like) Minkowski space signature:
\begin{eqnarray}
\det X=\xi^2-x_{n-1}^2-xx^*, \quad xx^*=x^*x=(x^0)^2+(x^1)^2+ ... +(x^{n-2})^2
\label{3.9}
\end{eqnarray}
and is thus invariant under the Lorentz group $\rm{SO}(n, 1)$. We leave it to the reader to verify that the multiplication law (\ref{3.7}) can be obtained from Eq. (\ref{1.5}) for the matrix (\ref{3.8}) by polarization. (Cf. Sect. 1.2.)

On the other hand, the algebras $\mathcal{H}_n(\mathbb{O})$ for $n>3$  are not Jordan since they violate condition $(ii)$ of (\ref{3.2}). The exceptional Jordan algebra $\mathcal{J}= \mathcal{H}_3(\mathbb{O})$ did not seem to be special but the authors of \cite{JvNW}  assigned the proof that the product $A\circ B$ of two elements of $\mathcal{J}$ cannot be represented in the form (\ref{3.1}) with an \textit{associative} product $AB$ to A. Adrian Albert (1905-1972) - a PhD student of L. Dickson.
As a result, many authors, including \cite{McC} call $\mathcal{J}$ an \textit{Albert algebra}.

With the realization (\ref{3.8}) of the elements of $JSpin_n$ we see that each (simple) euclidean Jordan algebra is a matrix algebra of some kind and so has a well defined (matrix) trace. The trace $\rho$ of the unit element of a Jordan algebra defines its \textit{rank}; in particular, the rank of the algerbra $JSpin_n$ for any $n$ is $\rho(JSpin_n)=2$. The real dimension
$\delta$ of the non-diagonal elements of the matrix represerntation of a Jordan algebra is called its \textit{degree}. (For a concise survey of euclidean Jordan algebras and their two numerical characteristics, the rank and the degree - see \cite{M}
(Sect. 2); note that Meng denotes the $JSpin_n$ by $\Gamma(n)$ and calls them \textit{Dirac type}.) The algebras listed in (\ref{3.5}) have degrees $\delta= 1, 2, 4, n-1$, respectively, while the Albert algebra (\ref{3.6}) has degree 8. The dimension of an euclidean Jordan algebra $V=J_{\rho}^\delta$ of rank $\rho$ and degree $\delta$ is $dim J_{\rho}^{\delta} = {\rho \choose 2} \delta + \rho$. The rank and the degree completely characterize the simple euclidean Jordan algebras.

We introduce the $1$-dimensional projectors $E_i$ and the hermitian octonionic matrices $F_i(x_i)$ writing down a general element of $\mathcal{H}_3(\mathbb{O})$ as
\begin{eqnarray}
X&=&\left( \begin{array}{ccc}
\xi_1  &  x_3  &  x_2^* \\
x_3^* &  \xi_2 & x_1 \\
x_2 & x_1^*  & \xi_3 \\
\end{array} \right) \nonumber \\
& =&\xi_1E_1+\xi_2E_2+\xi_3E_3+F_1(x_1)+F_2(x_2)+F_3(x_3) .
\label{3.10}
\end{eqnarray}
We can then write the Jordan multiplication $X\circ Y$ setting
\begin{eqnarray}
E_i \circ E_j&=&\delta_{ij}E_j, \ E_i\circ F_j(x)=
\begin{cases}
 0,\ \text{if} \ i=j, \\
 \frac{1}{2}F_j(x),\ \text{if} \ i\neq j. \nonumber \\
\end{cases}
\end{eqnarray}
\begin{eqnarray}
F_i(x)\circ F_i(y)&=&<x,y> (E_{i+1}+E_{i+2}), \nonumber \\ F_i(x)\circ F_{i+1}(y)&=&\frac{1}{2}F_{i+2}(y^*x^*),
\label{3.11}
\end{eqnarray}
where the indices are counted $\mod 3$: $E_4\equiv E_1$, \ $F_5\equiv F_2$, \dots . We define the trace, a symmetric bilinear inner product and a trilinear scalar product in $J$ by
\begin{eqnarray}
\rm{tr}X&= &\xi_1+\xi_2+\xi_3, \nonumber \\
<X,Y>&=& \rm{tr}(X\circ Y), \ \rm{tr}(X,Y,Z)=<X,Y\circ Z>.
\label{3.12}
\end{eqnarray}
The exceptional algebra $\mathcal{J}$ also admits a (symmetric) \textit{Freudenthal product}:
\begin{eqnarray}
X\times Y=\frac{1}{2}\left[2X\circ Y-X\rm{tr}Y-Y\rm{tr}X+(\rm{tr}X\rm{tr}Y-<X,Y>)E\right]
\label{3.13}
\end{eqnarray}
where $E$ is the $3\times 3$ unit matrix, $E=E_1+E_2+E_3$. Finally, we define a $3$-linear form $(X,Y,Z)$ and the determinant $\det X$ by
\begin{eqnarray}
(X,Y,Z)&=&<X,Y\times Z>= <X\times Y,Z>, \  \det X=\frac{1}{3}(X,X,X)
 \nonumber \\
= \xi_1\xi_2\xi_3&+& 2Re(x_1x_2x_3)-\xi_1x_1x_1^*-\xi_2x_2x_2^*-\xi_3x_3x_3^*.
\label{3.14}
\end{eqnarray}
The following identities hold:
\begin{eqnarray}
X\times X\circ X=(\det X)E \ \  (\text{Hamilton-Cayley}) \nonumber \\
(X\times X)\times(X\times X)=(\det X)X.
\label{3.15}
\end{eqnarray}
\bigskip
\subsection{The Tits-Kantor-Koecher (TKK) construction}
The symmetrized product $u\circ x$ (\ref{3.1}) is not the only way to construct a hermitean operator out of two such operators
$u$ and $x$. The quadratic in $u$ binary operation $P_ux =uxu$ also gives a hermitean result whenever $u$ and $x$ are hermitean. Its expression in terms of the Jordan product looks somewhat clumsy:
\begin{equation}
P_u=2L_u^2-L_{u^2},\quad - \text{i.e.} \quad P_ux = u\circ (u\circ x) - u^2\circ x
\label{P-L}
\end{equation}
but, as emphasized by McCrimmon \cite{McC}, it can be advantageously taken as a basic notion defined axiomatically.
Introduce first the polarized form of $P$:
\begin{equation}
S_{uv}w =\frac{1}{2}(P_{u+w} - P_u - P_w)v =  \{uvw\}:= u\circ (v\circ w) + w\circ (v\circ u)-(u\circ w)\circ v = \{wvu\}.
\label{Suv}
\end{equation}
A \textit{unital quadratic Jordan algebra} is a space together with a distinguished element 1 and a product $P_u(x)$ linear in
$x$ and quadratic in $u$, which is \textit{unital} and satisfies the \textit{Commuting Formula} and the \textit{Fundamental Formula}:
\begin{equation}
P_1=1, \qquad P_u S_{vu} = S_{uv} P_u, \quad P_{P_u(v)} = P_u P_v P_u.
\label{PSaxioms}
\end{equation}
The triple product $\{uvw\}$ is symmetric (according to the last equation (\ref{Suv})) and obeys the \textit{5-linear relation}
\begin{equation}
\{x y \{u v w\}\} = \{\{x y u\} v w\} - \{u \{y x v\} w\} + \{u v \{x y w\}\}.
\label{triple5}
\end{equation}
This identity can be read as a Lie algebra relation:
\begin{equation}
[S_{xy}, S_{uv}] = S_{\{x y u\} v} - S_{u\{y x v\}}
\label{str}
\end{equation}
applied to an arbitrary element $w\in V$. It defines the \textit{structure  Lie algebra} $\frak{str}$ of $V$. The generators
$S_{uv}$ can be expressed in terms of the (left) multiplication operators $L_x$ as follows:
\begin{equation}
S_u(=S_{u 1}) = L_u, \quad S_{uv} = L_{uv} + [L_u, L_v].
\end{equation}
The \textit{derivation algebra} $\frak{der}(V)$ of $V$, spanned by the commutators $\left[L_u,L_v\right]$, appears as the maximal compact Lie subalgebra of $\frak{str}(V)$.

The \textit{conformal algebra} $\frak{co}(v)$ is an extension of $\frak{str}(V)$
defined, as a vector space, as
\begin{eqnarray}
\frak{co}(v)=V\dotplus \frak{str}(V)\dotplus V^*
\label{co(V)}
\end{eqnarray}
with the natural \textit{TKK commutation relations} (under the action of the structure algebra, $u$ and $v$ in $S_{uv}$ transforming as a vector and a covector, respectively - see Theorem 3.1 of \cite{M}). Here there are three relevant examples
(the reader can find a complete list in \cite{M}):
\begin{eqnarray}
\frak{der}(JSpin_n)&=&\frak{so}(n),\quad \frak{str}(JSpin_n)=\frak{so}(n,1)\oplus\mathbb{R},\quad \frak{co}(JSpin_n)=\frak{so}(n+1,2)\nonumber \\
\frak{der}(\mathcal{H}_n(\mathbb{C}))& =&\frak{su}(n), \quad \frak{str}(\mathcal{H}_n(\mathbb{C}))=\frak{sl}(n,\mathbb{C})\oplus\mathbb{R},
\quad \frak{co}(\mathcal{H}_n(\mathbb{C}))=\frak{su}(n,n)\nonumber \\
\frak{der}(\mathcal{H}_3(\mathbb{O}))& =&\frak{f}_4, \qquad \frak{str}(\mathcal{H}_3(\mathbb{O}))=\frak{e}_{6(-26)}\oplus\mathbb{R},
\quad \frak{co}(\mathcal{H}_3(\mathbb{O}))=\frak\frak{e}_{7(-25)}.
\label{co-E7}
\end{eqnarray}

A \textit{Jordan triple system} V with a triple product $V^{\times 3}\rightarrow V, \{u v w \}$, satisfying (i) $\{u v w\} = \{w v u\}$ and (\ref{triple5}) is a generalization of a Jordan algebra (as every Jordan algebra generates a Jordan triple system). The same structure arises\footnote{This fact has been discovered by Isaiah Kantor (1964) - see the emotional essay \cite{Z} by Efim Zelmanov.} in any 3-graded Lie algebra $\frak{g}= \frak{g_{-1}}\dotplus \frak{g}_0 \dotplus\frak{g}_1$ with an involution $\tau$ exchanging $\frak{g}_{\pm 1}$ (see \cite{P}):
\begin{equation}
\{u v w\} = [[u, \tau(v)], w], \, u, v, w \in \frak{g}_1, \, \tau(v)\in  \frak{g}_{-1}, \, [u, \tau(v)]\in \frak{g}_0.
\end{equation}

Another important discovery, due to Koecher and his school, is the existence of a one-to-one correspondence between (simple) euclidean Jordan algebras $V$ and (irreducible) symmetric cones $\Omega(V)$ (see\cite{K}, \cite{FK}). To the four matrix algebras correspond the cones of positive definite  matrices $\Omega_n(\mathbb{R})$, $\Omega_n(\mathbb{C})$, $\Omega_n(\mathbb{H})$, $\Omega_3(\mathbb{O})$ of rank $\rho:=tr_{\vee}(1)$ equal to $n$ or $3$, respectively. The positive cone of $JSpin_n$ coincides with the forward light cone:
\begin{eqnarray}
\Omega_n(JSpin_n) =\{(\xi,x)\in JSpin_n| \xi>\sqrt{x_1^2+\dots+x_n^2}\}.
\label{OmegaJSpin}
\end{eqnarray}
In all cases $\Omega(V)$ is spanned by (convex) linear combinations of squares of elements $x\in V$ with positive coefficients; equivalently, $\Omega(V)$ is the connected component of the unit element $1\in V$ of the invertible elements of $V$. The cones $\Omega(V)$ are all selfdual and invariant under the structure group $Str(V)$. The conformal group $Co(V)$  can be defined as the automorphism group of the tube domain:
\begin{eqnarray}
Co(V)=Aut\{V+i \Omega(V)\}.
\label{Co(tube)}
\end{eqnarray}

\subsection{Automorphism groups of the exceptional Jordan\\ algebras $\mathcal {H}_3(\mathbb{O}_{(s)})$ and their maximal subgroups}
Classical Lie groups appear as symmetries of classical symmetric spaces.   For quite some time there was no such interpretation for  the exceptional Lie groups. The situation only changed with the discovery of the exceptional Jordan algebra $\mathcal {H}_3(\mathbb{O})$   and its split octonions' cousin
$\mathcal {H}_3(\mathbb{O}_s)$.

The automorphism group of  $\mathcal {H}_3(\mathbb{O})$ is the rank four compact simple Lie group \footnote{This was proven by Claude Chevalley and Richard Schafer in 1950. The result was prepared by Ruth Moufang's study in 1933 of the octonionic projective plane, then Jordan's construction in 1949 of $\mathbb{O}\mathbb{P}^2$ in terms of $1$-dimensional projections in $\mathcal {H}_3(\mathbb{O})$ and Armand Borel's observation that $F_4$ is the isometry group of $\mathbb{O}\mathbb{P}^2$; for a review and references - see \cite{B} (Sect. 4.2). Octonionic quantum mechanics in the Moufang plane was considered in \cite{GPR}.}  $F_4$. It leaves the unit element E invariant and is proven to preserve the trace (\ref{3.12}) (see Lemma 2.2.1 in \cite{Y}). The stabilizer of $E_1$ in $F_4$ is the double cover $Spin(9)$ of the rotation group in nine dimensions (which preserves $X_0^2$ (\ref{3.8})). Moreover, we have
\begin{eqnarray}
F_4/Spin(9) &\simeq& \mathbb{O}\mathbb{P}^2 \Longrightarrow
 \nonumber \\
\dim F_4&=&\dim Spin(9)+\dim \mathbb{O}^2=36+16=52.
\label{3.16}
\end{eqnarray}

Building on our treatment of $D_4\supset{\mathfrak g}_{2}$ of Sect. 2.4 we shall first construct the Cartan subalgebra of the Lie algebra $\frak{f}_4$ of \textit{derivation} (infinitesimal automorphisms)  of  $\mathcal{H}_3(\mathbb{O}_s)$. It is again spanned by the orthonormal weight basis $\Lambda_{\mu}$, $\mu=0,1,2,4$ (that actually belongs to the real form $f_{4(4)}\supset \frak{so}(4,4)$), their restriction to $\frak{D}_4$ being given by (\ref{2.41}).
The simple roots $\alpha$ and the corresponding coroots $\alpha^{\vee}$ of $\frak{f}_4$ are given by:
\begin{eqnarray}
\alpha_1&=&\Lambda_1-\Lambda_2=\alpha_1^{\vee}=H_1, \qquad
\alpha_2=\Lambda_2-\Lambda_4= \alpha_2^{\vee}=H_{2} \nonumber \\
(\alpha_4^{(s)}=)s_4&=&\Lambda_4, \ s_4^{\vee}=2\Lambda_4=H_4^s,\quad s_0=\frac{1}{2}(\Lambda_0-\Lambda_1-\Lambda_2-\Lambda_4)(=\alpha_0^{(s)})\nonumber \\
s_0^{\vee}&=&\Lambda_0-\Lambda_1 -\Lambda_2-\Lambda_4=H_0^s.
\label{3.17}
\end{eqnarray}
The corresponding Cartan matrix reads:
\begin{eqnarray}
 (c^f_{ij}=<\alpha_i^\vee,\alpha_j>) = \left( \begin{array}{cccc}
 2 & -1 &  0 &  0 \\
-1 &  2 & -2 &  0 \\
 0 & -1 &  2 & -1 \\
 0 &  0 & -1 &  2 \\
\end{array} \right).
\label{3.18}
\end{eqnarray}
%The Lie algebra $\mathfrak f_{4(4)}$ has $24$ positive roots: the $12$ long roots coincide with the positive roots \{$\lambda_0\pm \lambda_j,\ j=1,2,4,\ \lambda_j\pm \lambda_k,\ 1\leqq j < k \leqq 4$\} of $so(4,4)$; the $4$ short roots $\lambda_{\mu},\ \mu=0,1,2,4$ coincide with the short positive roots of $so(5,4)$; finally, $\mathfrak f_{4(4)}$ has $8$ additional short roots of the form $\frac{1}{2}(\lambda_0\pm\lambda_1,\pm\lambda_2\pm\lambda_4)$;
The highest root $\theta $ coincides with that of the rank four simple subalgebra   $\mathfrak B_{4} \supset\frak{D}_4$ (respectively, of the real forms  $so(5,4) \supset so(4,4)$):
\begin{eqnarray}
 \theta =\Lambda_0+\Lambda_1(=2\alpha_1+3\alpha_2+4\alpha_4+2\alpha_0).
\label{3.19}
\end{eqnarray}
The elements $D$ of $so(8)$ act on $X$ of $\mathcal{J}$ (\ref{3.10}) through their action on the octonions.
\begin{eqnarray}
DX=F_1(Dx_1)+F_2(\kappa(D)x_2)+F_3(\pi (D)x_3),
\label{3.20}
\end{eqnarray}
where $D=:D_1$, $\kappa(D)=:D_2$, $\pi(D)=:D_3$, obey the principle of infinitesimal triality:
\begin{eqnarray}
(D_1x)y+x(D_2y)=(D_3((xy)^*))^*.
\label{3.21}
\end{eqnarray}
For $D\in G_2$ we have $D_1=D_2=D_3=D$.

The remaining $24$ generators of $\mathfrak f_4$ (outside $so(8)$) can be identified with the skew-hermitian matrices $A_i(e_a)$, $i=1,2,3$, $a=0,1,\dots, 7$
\begin{eqnarray}
 A_1(x)&=& \left( \begin{array}{ccc}
0 & 0  &0  \\
0  & 0 & x \\
0  &-x^* &  0 \\
\end{array} \right), \, \,
A_2(x)= \left( \begin{array}{ccc}
0 & 0  & -x^* \\
0 & 0 &  0   \\
x & 0 &  0  \\
\end{array} \right), \nonumber \\
A_3(x)&=& \left( \begin{array}{ccc}
0 & x & 0  \\
-x^*& 0 & 0 \\
0 & 0 &  0  \\
\end{array} \right).
\label{3.22}
\end{eqnarray}
They act on $\mathcal{J}$ through the commutators
\begin{eqnarray}
\tilde{A_i}(e_a)X=\frac{1}{2}[A_i(e_a),X],\ i=1,2.3,\ a=0,1,...,7.
\label{3.23}
\end{eqnarray}

The Borel - de Siebenthal theory \cite{BdS} describes the maximal rank closed connected subgroups of a compact Lie group. In order to reveal the physical meaning of the symmetry of $\mathcal{H}_3(\mathbb{O})$ we shall consider, elaborating on \cite{TD}, those maximal rank subgroups of $\rm{F}_4$ which contain the (unbroken) colour symmetry group $\rm{SU}(3)_c\subset \rm{G}_2\subset\rm{F}_4$. There are two such subgroups:
\begin{eqnarray}
\frac{SU(3) \times SU(3)}{\mathbb{Z}_3} \qquad \text{and} \qquad \rm{Spin}(9).
\label{3.24}
\end{eqnarray}
Postponing the study of
the maximal subgroup $\rm{Spin}(9)$ to Sect. 4.1 we shall display here the action of

\begin{eqnarray}
F_4^{\omega}=\frac{SU(3) \times SU(3)}{\mathbb{Z}_3}, \quad \omega(a+z^je_j)=a+\omega_7z^Je_j \quad (\omega^3=1=\omega_7^3)
\label{3.25}
\end{eqnarray}
\begin{eqnarray}
(a=a^0+a^7e_7, \quad z^je_j=z^1e_1+z^2e_2+z^4e_4 )\nonumber
\end{eqnarray}
(cf. (\ref{2.32})) on the exceptional Jordan algebra.
 To do that we shall first extend the splitting of the octonions $\mathbb{O}=\mathbb{C} \oplus \mathbb{C}^3$ to a splitting of the exceptional Jordan algebra, $\mathcal{H}_3(\mathbb{O})=\mathcal{H}_3(\mathbb{C}) \oplus \mathbb{C}[3]$:
\begin{eqnarray}
(\mathcal{H}_3(\mathbb{O})\ni) X(\xi,x)= \left( \begin{array}{ccc}
\xi_1   & x_3    &  x_2^* \\
x_3^*  & \xi_2  &  x_1   \\
x_2    & x_1^*  &  \xi_3 \\
\end{array} \right)= X(\xi,a)+X(0,\textbf{z}\textbf{e})
\label{3.26}
\end{eqnarray}
where
\begin{eqnarray}
  X(\xi,a)&=& \left( \begin{array}{ccc}
\xi_1           & a_3             &  \overline{a}_2 \\
\overline{a}_3  & \xi_2           & a_1   \\
a_2             & \overline{a}_1  & \xi_3 \\

\end{array} \right),  \nonumber \\
  a_r&=& x_r^0+ x_r^7e_7, \ \overline{a}_r=x_r^0-x_r^7e_7,\ r=1,2,3;
 \nonumber \\
 X(0,\mathbf{z}\mathbf{e})&=& \left( \begin{array}{ccc}
0   & \mathbf{z}_3\mathbf{e}  & - \mathbf{z_2}\mathbf{e}  \\
- \mathbf{z_3}\mathbf{e} & 0& \mathbf{z_1}\mathbf{e} \\
\mathbf{z_2}\mathbf{e} & -\mathbf{z_1}\mathbf{e} & 0   \\
\end{array} \right),  \nonumber \\
\mathbf{z_r}\mathbf{e}&=& z_r^1e_1+z_r^2 e_2+z_r^4 e_4, z_r^j=x_r^j+x_r^{3j(mod 7)}e_7 ,
\label{3.27}
\end{eqnarray}
(we have used the conjugation property $(\textbf{z}\textbf{e})^*=-\textbf{z}\textbf{e}$ of imaginary octonions). Multiplication mixes the two terms in the right hand side of (\ref{3.26}). The Freudenthal product $X(\xi,x)\times Y(\eta,b)$ can be expressed in a nice compact way if we substitute the skew symmetric octonionic matrices $X(0,\textbf{z}\textbf{e}), X(0,\textbf{w}\textbf{e})$   by  $3\times 3$ complex matrices $Z, W$:
\begin{eqnarray}
X(0,\mathbf{z}\mathbf{e})\longleftrightarrow Z=(z_r^j,\ r=1,2,3, \, \, \jmath=1,2,4) \in \mathbb{C}[3],
\label{3.28}
\end{eqnarray}
which transform naturally under the subgroup (\ref{3.25}).

Indeed, using the fact that the matrices $X(0,\textbf{z}\textbf{e})$ and $X(0,\textbf{w}\textbf{e})$ are traceless, their Fredenthal product $( \ref{3.13})$ simplifies and we find:
\begin{eqnarray}
&& X(\xi,a)\times X(0,\mathbf{w}\mathbf{e})= X(\xi,a)\circ  X(0,\mathbf{w}\mathbf{e})- \frac{\xi_1+\xi_2+\xi_3}{2}X(0,\mathbf{w}\mathbf{e}) \nonumber  \\ 	 &&\Longrightarrow X(\xi,a)\times W=-\frac{1}{2}WX(\xi,a), \ \text{for} \ W=(w^j_{ r});
\label{3.29}
\end{eqnarray}
\begin{eqnarray}
X(0,\mathbf{z}\mathbf{e})\times X(0,\mathbf{w}\mathbf{e})&=& X(0,\mathbf{z}\mathbf{e})\circ X(0,\mathbf{w}\mathbf{e})-\frac{1}{2}tr(X(0,\mathbf{z}\mathbf{e})X(0,\mathbf{w}\mathbf{e}))E \nonumber \\
X(0,\mathbf{z}\mathbf{e})\times X(0,\mathbf{w}\mathbf{e}) &\leftrightarrow & -\frac{1}{2} (W^*Z+Z^*W +\overline{Z}\times\overline{W})
\label{3.30}
\end{eqnarray}
where $Z\times W=(\epsilon_{rst}(\textbf{z}_s \times \mathbf{w}_t)^j))$, so that
\begin{eqnarray}
(X(\xi,a)+Z)\times(X(\eta,b)+W)=X(\zeta,c)+V  \nonumber  \\
X\zeta,c)=X(\xi,a) \times X(\eta,b)-\frac{1}{2}(Z^*W-W^*Z)  \nonumber \\
V=-\frac{1}{2}\left(WX(\xi,a)+ZX(\eta,b) +\overline{Z}\times \overline{W}\right).
\label{3.31}
\end{eqnarray}
Thus, if we set $V=(v^j_r)$ we would have
\begin{eqnarray}
2\mathbf{v_1}=-\xi_1\mathbf{w_1} -\overline{a}_3\mathbf{w_2}-a_2\mathbf{w_3}-\mathbf{\overline {z}_2}\times\mathbf{\overline{ w}_3}  \nonumber \\
2\mathbf{v_2}=-a_3\mathbf{w_1} -\xi_2\mathbf{w_2}-\overline{ a}_1\mathbf{w_3}-\mathbf{\overline{z}_3}\times\mathbf{\overline{w}_1}  \nonumber \\
2\mathbf{v_3}=-\overline{a}_2\mathbf{w}_1 -a_1\mathbf{w}_2-\xi_3\mathbf{w}_3-\mathbf{\overline {z}_1} \times\mathbf{\overline{ w}_2}.
\nonumber
\end{eqnarray}
The inner product in $\mathcal{J}$ is expressed in terms of the components $X(\xi,a)$ and $Z$ as:
\begin{eqnarray}
(X,Y)(=trX\circ Y)=(X(\xi,a),X(\eta,b))+2(Z,W) \nonumber \\
\text{where} \, \, 2(Z,W)=Tr(Z^*W+W^*Z)=2\sum_{r=1}^3\sum_{j=1,2,4}(\overline {z}_r^jw_r^j+\overline{w}_r^jz_r^j).
\label{3.32}
\end{eqnarray}
In the applications to the standard model of particle physics the (upper) index $j$ of $z$ ($j=1,2,4$) labels quark's colour while $r\in \{1,2,3\}$ is a \textit{flavour} index. The $SU(3)$ subgroup of $G_2$, displayed in Sect. 2 acting on individual (imaginary) octonions is the colour group.

The subgroup $F_4^{\omega}$ (\ref{3.25}) is defined as the  commutant of the automorphism $\omega$ of order three in $F_4$ (see (\ref{2.32})):
\begin{eqnarray}
 \omega X(\xi,x)&=& \left( \begin{array}{ccc}
\xi_1 & \omega x_3 & (\omega x_2)^*  \\
(\omega x_3)^*  & \xi_2 & \omega x_1  \\
\omega x_2  & (\omega x_1)^*&  \xi_3 \\
\end{array} \right), \nonumber \\ %\omega(a&+&\mathbf{z}\mathbf{e})=a+\omega_7\mathbf{z}\mathbf{e}, \nonumber \\
%\omega_7&=&-\frac{1}{2}+\frac{\sqrt3}{2}e_7,\ (\omega_7^3=1=\omega^3),
%\nonumber \\
\omega(X(\xi,a)+Z)&=&X(\xi,a)+\omega_7Z.
\label{3.33}
\end{eqnarray}
 The automorphisms $g\in F_4^{\omega}$ that commute with $\omega$(\ref{3.33}) are given by pairs $g=(A,U)\in SU(3)\times SU(3)$ acting on $\mathcal{H}_3(\mathbb{O})$ by
\begin{eqnarray}
(A,U)\left(X(\xi,a)+Z\right)=AX(\xi,a)A^*+UZA^*.
\label{3.34}
\end{eqnarray}
The central subgroup
\begin{eqnarray}
\mathbb{Z}_3=\{(1,1),(\omega_7,\omega_7),(\omega_7^2,\omega_7^2)\}\in
 SU(3)\times SU(3)
\label{3.35}
\end{eqnarray}
acts trivially on $\mathcal{H}_3(\mathbb{O})$. We see that the unitary matrix $U$ acts (in (\ref{3.33})) on the colour index $j$ and hence belongs to the (unbroken) \textit{colour group} $SU_c(3)$, while the action of $A$ on the flavour indices will be made clear in Sect. 4.2 below.
%spans the (badly broken) family symmetry.

\smallskip

\section{$\rm{F}_4$ as a grand unified symmetry\\ of the standard model}
\setcounter{equation}{0}
\renewcommand\theequation{\thesection.\arabic{equation}}

This chapter provides a tentative application of the exceptional Jordan algebra to the standard model (SM) of particle physics. We first study, in Sect. 4.1, the special Jordan subalgebra $JSpin_9$ of $\mathcal{J}$ and its automorphism group $Spin(9)\subset\rm{F}_4$  singling out the $16$-dimensiional spinor representation of $Spin(9)$ and interpret it in terms of the fundamental fermionic doublets of the SM. Then, in Sect.4.2, we examine the full $26$-dimensional representation \underline{26} of $\rm{F}_4$, considering its restriction to both its maximal rank subgroups that contain the colour $\rm{SU}(3)_c$ as subgroup. We demonstrate that \underline{26} gives also room, to the electroweak gauge bosons.

\subsection{The Jordan subalgebra $JSpin_9$ of $\mathcal{H}_3(\mathbb{O})$ }
The ten dimensional Jordan algebra $JSpin_9$ can be identified with the algebra of $2\times2$ hermitian octonionic matrices $\mathcal{H}_2(\mathbb{O})$ equipped with the Jordan matrix product. It is generated by the $9$-dimensional vector subspace
$s\mathcal{H}_2(\mathbb{O})$ of traceless matrices of $\mathcal{H}_2(\mathbb{O})$ whose square is, in fact, a positive real scalar:
\begin{eqnarray}
	X= \left( \begin{array}{cc}
		\xi   & x     \\
		x^*   &-\xi   \\
	\end{array} \right)\Rightarrow  X ^2 = (\xi^2+x^*x)1, \nonumber \\
	x\in \mathbb{O}, \  \xi \in\mathbb{R}.
	\label{4.1}
\end{eqnarray}
$JSpin_9$ is a (special) Jordan subalgebra of the (associative) matrix algebra $\mathbb{R}[2^4]$ that provides an irreducible representation of $C\ell_9$. Clearly, it is a subalgebra of $\mathcal{H}_3(\mathbb{O})$-consisting of $3\times3$ matrices with vanishing first row and first column. Its automorphism group is the subgroup $\rm{Spin}(9)\subset F_4$ which stabilizes the projector $E_1$: $\rm{Spin}(9)=(F_4)_{E_1}\subset F_4$.

We shall see that the spinor rpresentation of $Spin_9$ can be interpreted as displaying the first generation of (left chiral) doublets of quarks and leptons
\[ \left( \begin{array}{cc}
\nu_L   & u^j_L     \\
e^-_L   & d^j_L   \\
\end{array} \right)(j  \ \text{ is the colour index}) \]
and their antiparticles. As demonstrated in \cite{D-VT} and elaborated in Sect. 4.2 below (Eqs.(4.17-21) the exceptional Jordan algebra gives room to a similar construction of the second and third generation fermions.

The $16$-dimensional (real) spinor representation $S_9$ of $\rm{Spin}(9)$ splits into a direct sum of the two $8$-dimensional chiral representations $S_8^+$ and $S_8^-$ of Spin(8) that appear as eigenvectors of the Coxeter element $\omega_8$ of $Cl_8$:
\begin{eqnarray}
	S_9=S_8^+ \oplus S_8^-,\qquad  \omega_8 S_8^\pm=\pm S_8^\pm.
	\label{4.2}
\end{eqnarray}
We shall use the relation (\ref{2.38}) ($\hat{P}_0=\sigma_1\otimes P_0, \ \hat{P}_a=c\otimes P_a,\ \ a=1,\dots, 7$) with real $P_j$ ($j=1,2,4$) and $i P_7, i P_3,  i P_6, i P_5$.
\begin{eqnarray}
	P_1 &=&\un\otimes\sigma_1\otimes c, \quad   \ P_2 =\sigma_1\otimes \sigma_3 \otimes c^* , \quad  P_4= c\otimes\sigma_1 \otimes  \sigma_1  \nonumber  \\
	i P_7 &=&\un \otimes \un \otimes \sigma_3, \quad i P_3=\sigma_3\otimes \sigma_1\otimes \sigma_1, \quad  i P_6= -\sigma_1\otimes \un \otimes \sigma_1,  \nonumber \\
	i P_5&=&\sigma_1 \otimes c\otimes c, \quad \Rightarrow \quad \omega_{-7}:=P_1P_2P_3P_4P_5P_6P_7=-\un\otimes\un\otimes\un,
   	\label{4.3}
\end{eqnarray}
\begin{eqnarray}
	\un=\left(\begin{array}{cc}
		1  & 0  \\
		0  & 1  \\
	\end{array}\right),\quad c=i\sigma_2,\quad c^*=-c=\left(\begin{array}{cc}
	0 & -1  \\
	1 &  0  \\
\end{array}\right). \nonumber
\end{eqnarray}
The Coxeter element $\omega_8$ of $C\ell_8$, appearing in (\ref{4.2}) is given by:
\begin{eqnarray}
	\hat{P}_8=\omega_8=\hat{P}_0\hat{P}_1\dots\hat{P}_7=\sigma_3\otimes\omega_{-7}=-\sigma_3\otimes\un\otimes\un\otimes \un .
	\label{4.4}
\end{eqnarray}
Inserting the matrices in the  tensor products from right to left, so that
\begin{eqnarray}
	\un\otimes\sigma_3= \left(
	\begin{array}{cc}
		\sigma_3 &  0      \\
		0    & \sigma_3\\
	\end{array}\right), \
	\sigma_3\otimes\un=\left(
	\begin{array}{cc}
		\un&  0      \\
		0&  -\un     \\
	\end{array}\right), \
	\sigma_3\otimes \sigma_3= \left(
	\begin{array}{cc}
		\sigma_3&  0      \\
		0&  -\sigma_3     \\
	\end{array}\right)
	\nonumber
\end{eqnarray}
we can write the weight matrices $\Lambda_{\mu}$ (\ref{2.41}) as:
\begin{eqnarray}
\Lambda_0&=&\frac{i}{2}\hat{P}_{07}=-\frac{1}{2}\sigma_3\otimes \un \otimes \un \otimes \sigma_3=-\frac{1}{2}{\rm diag} \left(\sigma_3,\sigma_3,\sigma_3,\sigma_3,-\sigma_3,-\sigma_3,-\sigma_3,-\sigma_3\right)\nonumber \\
\Lambda_1&=&\frac{i}{2}\hat{P}_{13}=-\frac{1}{2}\un\otimes \sigma_3\otimes \un \otimes \sigma_3=-\frac{1}{2}\un\otimes {\rm diag} \left(\sigma_3,\sigma_3,-\sigma_3,-\sigma_3\right) \nonumber \\
\Lambda_2&=&\frac{i}{2}\hat{P}_{26}=-\frac{1}{2} \un\otimes \un \otimes \sigma_3\otimes \sigma_3=-\frac{1}{2}\un\otimes {\rm diag} \left(\sigma_3,-\sigma_3,\sigma_3,-\sigma_3\right) \nonumber \\
\Lambda_4&=&\frac{i}{2}\hat{P}_{45}=-\frac{1}{2} \un \otimes \sigma_3\otimes \sigma_3 \otimes \sigma_3=-\frac{1}{2}\un\otimes {\rm diag} \left(\sigma_3,-\sigma_3,-\sigma_3,\sigma_3\right).
\label{4.5}
\end{eqnarray}
They form a (commuting) orthonormal basis in the weight space:
\begin{eqnarray}
\Lambda_{\mu}^2=\frac{1}{4}1_{16},\ [\Lambda_{\mu},\Lambda_{\nu}]=0, \ <\Lambda_{\mu},\Lambda_{\nu}>:=\frac{1}{4}tr(\Lambda_{\mu}\Lambda_{\nu})=\delta_{\mu\nu}, \mu,\nu=0,1,2,4. \nonumber
\end{eqnarray}
The Lie algebra $\frak{B}_4(=\frak{so}(9)$) of $\rm{Spin}(9)$ is spanned by the commutators $\hat{P}_{ab}=\frac{1}{2}[\hat{P}_a, \hat{P}_b],\ a,b=0,1, \dots,8$.
The physical meaning of $\frak {B}_4$ is best revealed by identifyimg its maximal (rank 4) subalgebra
\begin{eqnarray}
\frak{su}(2)\oplus\frak{su}(4)\approxeq\frak{so}(3)\oplus\frak{so}(6)\in\frak{so}(9).
\label{4.6}
\end{eqnarray}
that is part of the\textit{ Pati-Salam grand unified Lie algebra}
\begin{eqnarray}
\frak{g}_{PS}:=\frak{su}(2)_L\oplus\frak{su}(2)_R\oplus\frak{su}(4) .
\label{4.7}
\end{eqnarray}
(See  for a nice review \cite{BH}.) The $\rm{SU}(4)$ of Pati-Salam is designed to unify the quark colour group $\rm{SU}(3)_c$ with the lepton number. The colour Lie algebra $\frak{su}(3)_c$ is identified with the commutant in $\frak{su}(4)$ of the hypercharge $Y({\underline{4}})$. In the defining $4$-dimensional representation $\underline{4}$ of $\frak{su}(4)$ it is given by the traceless diagonal matrix
\begin{eqnarray}	
Y(\underline{4})= \left( \begin{array}{cccc}
\frac{1}{3}& 0 &  0 & 0 \\
0  &\frac{1}{3}&  0 & 0 \\
0  &  0 &\frac{1}{3}& 0 \\
0  &  0 &  0        &-1        \\
\end{array} \right).
\label{4.8}
\end{eqnarray}
With our splitting of the octonions $\mathbb{O}=\mathbb{C}+\mathbb{C}^3$
and the corresponding embedding of $\rm{SU}(3)_c$ into $\rm{G}_2$ (see (\ref{2.11}))
the subalgebra $\frak{so}(6)$ is spanned by $\hat{P}_{jk}$ with $1\leq j<k\leq 6$. The Lie subalgebra $\frak{su}(3)_c$ appears as the intersection of $\frak{g}_2$ and $\frak{so}(6)$ in $\frak{so}(7)$. As a consequence of (\ref{2.43}) it is spanned by the following combinations of $\hat{P}_{ab}$:
\begin{eqnarray}
	\frak{su}(3)_c = Span\{\hat{P}_{13}-\hat{P}_{26}(=-i H_1),\ \hat{P}_{26}-\hat{P}_{45}(-i H_2), \nonumber \\
	\hat{P}_{12}+\hat{P}_{36}, \qquad \hat{P}_{24}+\hat{P}_{45}, \qquad\hat{P}_{14}+\hat{P}_{35}  \nonumber \\
	\hat{P}_{16}+\hat{P}_{23}, \qquad \hat{P}_{25}+\hat{P}_{46}, \qquad \hat{P}_{15}+\hat{P}_{43}\}.
	\label{4.9}
\end{eqnarray}
The first two span the Cartan subalgebra of $\frak{su}(3)_c$ - cf. (\ref{3.17}). The next three span the maximal compact subalgebra $\frak{so}(3)$ of the real form $\frak sl(3,\mathbb{R})\subset\frak{so}(4,4)$. The generator $Y$ of the commutant of $\frak{su}(3)_c$ in (the complexifation of) $\frak{su}(4)$ is given by
\begin{eqnarray}
	Y=\frac{2}{3}\left(\Lambda_1+\Lambda_2+\Lambda_4\right)= \frac{i}{3} \left(\hat{P}_{13}+\hat{P}_{26}+\hat{P}_{45}\right)\quad (\in \frak{so}(3,3)),	
	\label{4.10}
\end{eqnarray}
\begin{eqnarray}	
3Y= \un_4\otimes \left( \begin{array}{cccc}
-3\sigma_3& 0 &  0 & 0 \\
0  &\sigma_3&  0 & 0 \\
0  &  0 &\sigma_3& 0 \\
0  &  0 &  0 &\sigma_3\\
\end{array} \right)=\un_4\otimes \rm {diag}(-3\sigma_3,\sigma_3,\sigma_3,\sigma_3) \nonumber.
%\label{4.16}
\end{eqnarray}

It (commutes with and) is orthogonal to the Cartan $\{H_1,H_2\}$ of $\frak{su}(3)_c$. The algebra $\frak{su}(2)$ in the left hand side of (\ref{4.6})  is then spanned by

\begin{eqnarray}
	I_3:=\Lambda_0=\frac{i}{2}\hat{P}_{07}, \qquad I_{+}=\hat{P}^{-}_{0}\hat{P}_8, \qquad I_{-}=\hat{P}_8\hat{P}^{+}_{0}  \nonumber \\
	(	\left[I_3, I_{\pm}\right]=\pm I_{\pm}, \qquad \left[I_{+},I_{-}\right]=2I_3).
	\label{4.11}
\end{eqnarray}
The spinor representation \underline{16} of ${\rm Spin(9)}$ can now be associated with the doublet representation
\begin{eqnarray}
	(\underline{4}^*,\underline{2},\underline{1})\oplus(\underline{4},\underline{2},\underline{1})\in \frak{g}_{PS}.
	\label{4.12}
\end{eqnarray}`
It is natural to identify $\Lambda_0$ which commutes with $\frak{su}(4)$ and has eigenvalues $\pm \frac{1}{2}$ with the third component $I_3$ of the weak isospin, while $Y$ that commutes with $\frak{su}(3)_c$  and is orthogonal to $I_3$ should coincide  with the weak hypercharge. The spinor representation consists of two octets of the fermion (lepton-quark) left chiral doublets and antifermion right chiral doublets
\begin{eqnarray}
	\left( \begin{array}{c}
		\nu_L       \\
		e^-_L      \\
	\end{array} \right),\quad Y=-1, \qquad
	\left( \begin{array}{c}
		u_L     \\
		d_L   \\
	\end{array}\right),\quad Y=\frac{1}{3}, \quad I_3=
	\left( \begin{array}{c}
		\frac{1}{2} \\
		-\frac{1}{2}\\
	\end{array} \right), \nonumber \\
	\left( \begin{array}{c}
		e^+_R       \\
		\overline{\nu}_R  \\
	\end{array} \right),\qquad Y=1, \quad
	\left( \begin{array}{c}
		\overline{d}_R  \\
		\overline{u}_R  \\
	\end{array}\right),\qquad Y=-\frac{1}{3}, \qquad I_3=
	\left( \begin{array}{c}
		\frac{1}{2} \\
		-\frac{1}{2}\\
	\end{array} \right).
	\label{4.13}
\end{eqnarray}
These are the fundamental (anti) fermions that participate in the weak interactions.
The missing fundamental fermion representations are the sixteen singlets with respect to the weak isospin, $I=0(=I_3$).
\subsection{The representation \underline{26} \, of $\rm{F}_4$}
As the automorphism group ${\rm F}_4$ of $\mathcal{J}=\mathcal {H}_3(\mathbb{O})$
preserves the unit element and the trace of $\mathcal{J}$ it acts faithfully and irreducibly on the $26$-dimensional subspace  $\mathcal{J}_0=s\mathcal {H}_3(\mathbb{O})\subset \mathcal{J}$ of traceless $3\times3$ hermitian octonionic matrices. In fact, \underline{26} is the lowest nontrivial (fundamental) representation (of highest weight $\Lambda_0$) of ${\rm F}_4$. Restricted to the maximal subgroup ${\rm Spin}(9)$ of ${\rm F}_4$ it splits into three irreducible components:
\begin{eqnarray}
	\underline{26}=\underline{16}+\underline{9}+\underline{1}.
	\label{4.14}
\end{eqnarray}
We have identified in the preceding Sect. 4.1 the spinor representation as four isospin chiral doublets of quarks and leptons and their antiparicles. The $9$-vector representation of $\frak{so}(9)$ is spanned by the generators $\hat{P}$ of $C\ell_9$. The matrices $\hat{P}^{\epsilon}_{\mu}$ (\ref{2.40}) and $\hat{P}_8$ diagonalize the adjoint action of the physical Cartan elements $Y$ (\ref{4.10}) and $I_3$ (\ref{4.11}); we find:
\begin{eqnarray}
	\left[I_3,\hat{P}_0^{\mp}\right]&=&\pm\hat{P}_0^{\mp}, \qquad \left[I_3,\hat{P}_8\right]=0=\left[I_3,\hat{P}^{\epsilon}_j \right],\quad (j=1,2,4) \nonumber \\
	\left[Y,\hat{P}_0^{\mp}\right]&=&0=\left[Y,\hat{P}_8\right], \qquad
	\left[Y,\hat{P}_j^{\mp}\right]=\pm\frac{2}{3}\hat{P}_j^{\mp}.
\label{4.15}
\end{eqnarray}
This allows to consider $\hat{P}_0^{\mp}$ as representatives of the $W^{\pm}$ bosons while $\hat{P}_8$ appears as the neutral component, $W^0$, of the isotopic triplet. Then the $Z$ boson and the photon can be associated with appropriate mixtures of $W^0$ and the trivial representation of $\frak{so}(9)$. The isotropic matrices $\hat{P}_j^{\pm}$, on the other hand, carry the quantum numbers ($I_3=0,\ Y=\mp\frac{2}{3}$) of the right handed $d$-quarks $d_R$ and  the left handed $d$-antiquarks,  $\overline{d}_L$, respectively ($j=1,2,4$ being the colour index). Let us note that all eigenvalues of the pair $(I_3,Y)$ appearing in the representation \underline{26} of $\rm{F}_4$ satisfy the SM constraint
$$I_3+\frac{3}{2}Y\in\mathbb Z. $$
We recall that according to the identification of the $2^5 $-dimensional fermion Fock space with the exterior algebra $\Lambda \mathbb{C}^5$ (see \cite{BH}) all fundamental fermions (of the first generation)  can be expressed as exterior products of $d_R, e_R^+$, and $\overline{\nu}_R$ (with $\overline{\nu}_L$ identified with the Fock vacuum). In particular, all fundamental fermions can be written as wedge products of $d_R$, $\overline{d}_L$ and the chiral spinors in (\ref{4.13}):
\begin{eqnarray}
	&& e_L^+(I_3=0,Y=2)= e_R^+\wedge \overline{\nu}_R \nonumber \\ &&u_R(I_3=0,Y=\frac{4}{3})=e_L^+\wedge d_R \nonumber \\
	&&\overline{u}_L^j(I_3=0,Y=-\frac{4}{3})=\epsilon^{jkl} d_R^k\wedge d_R^l\nonumber \\
	&& e_R^-(I_3 =0,Y=-2)=\sum_{j=1,2,4}\overline{u}_L^j\wedge d_R^j \nonumber \\
	&&\nu_R(I_3=0=Y)=\sum_{j}\overline{d}_L^j\wedge d_R^j, \quad
	(j,k,l=1,2,4) .
\label{4.16}
\end{eqnarray}
(Here $\epsilon^{ijk}$  is the fully antisymmetric unit tensor with $\epsilon^{124}=1$.) The isosinglets $\bar{\nu}_L, d_R, \bar{d}_L$ and the wedge products (\ref{4.16}) together with the isotopic doublets (\ref{4.13}) exhaust the 32 first generation fundamental fermions of the SM (cf. Sect. 5.3). We assume that right handed fermions have negative chirality so that wedge product with $d_R$ or $\overline{\nu}_R$ changes the sign of chirality (i.e. transforms right to left and vice versa). If we associate the Higgs boson with the trace (or the unit element) of the exceptional Jordan algebra $\mathcal{J}$  we see that all fundamental particles of the first family of the SM are generated  either directly or as exterior products of elements of $\mathcal{J}$.
The same assignment of quantum numbers to fundamental particles is obtained, as it should, if we consider the reduction of $\rm{F}_4$ with respect to the other admissible maximal subgroup $\rm{F}_4^{\omega}$ (\ref{3.25}). On the way of demonstrating this we shall express the third component of the weak isospin $I_3$ and the hypercharge $Y$ in terms of the Cartan elements of the flavour $\rm{SU}(3)$. 
%In the defining representation \underline{3} of the first  $\rm{SU}(3)$ %(implemented by $A$ in (\ref{3.34})); we set
%\begin{eqnarray}
%	2I_k(\underline{3})=\lambda_k:= \left( \begin{array}{cc}
%		\sigma_k & 0    \\
%		0  & 0  \\
%	\end{array} \right), \, \, k=1, 2, 3, \qquad 3Y(\underline{3})= \left( %\begin{array}{ccc}
%	-1 & 0  & 0 \\
%	0 & -1 & 0 \\
%	0 & 0 &  2 \\
%\end{array} \right) \nonumber \\
%\left( \sigma_1=\left( \begin{array}{cc}
%%	0 & 1   \\
%	1 & 0  \\
%\end{array} \right),\quad \sigma_2= \left( \begin{array}{cc}
%0 & -i  \\
%i & 0  \\
%\end{array} \right),\quad \sigma_3=\left( \begin{array}{cc}
%1 &  0   \\
%0  & -1 \\
%\end{array} \right)\right)
%\label{4.17}
%\end{eqnarray}
%($\lambda_a$, $a=1,2\dots,8$ are the Gell-Mann $\frak {su}(3)$ matrices). These operators act on the part $X(\xi,a)$ (\ref{3.27}) of $\mathcal{J}$ by commutation and on the matrix $Z$ (\ref{3.28}) by right multiplication with $(-I_3,-Y)$. Their eigenvectors are given by
To begin with we introduce the three Weyl units $\epsilon_i$ and their conjugate (=transposed) $\epsilon_i^*$ ($i=1,2,3$) by
\begin{eqnarray}
	\epsilon^*_1&=&\left( \begin{array}{ccc}
		0 &	0 & 0   \\
		0 & 0 & 1 \\
		0 & 0 & 0  \\
	\end{array} \right),
	\quad \epsilon^*_2= \left( \begin{array}{ccc}
		0 &	0 & 0  \\
		0 & 0 & 0  \\
		1 & 0 & 0  \\
	\end{array} \right),
	\quad \epsilon^*_3= \left( \begin{array}{ccc}
		0 &	1 & 0  \\
		0 & 0 & 0   \\
		0 & 0 & 0   \\
	\end{array} \right) 
%	Z_1&=&(z^j_1,0,0), \qquad Z_2=(0,z^j_2,0), \qquad Z_3=(0,0,z^j_3)
\label{4.17}
\end{eqnarray}
and set
\begin {eqnarray}
2I_3&=&[\epsilon^*_1,\epsilon_1]=
\left( \begin{array}{ccc}
	0 &	0 & 0 \\
	0 & 1 & 0 \\
	0 & 0 &-1 \\
\end{array} \right)\nonumber \\
3Y&=&[\epsilon^*_3,\epsilon_3]-[\epsilon^*_2,\epsilon_2]= 
\left( \begin{array}{ccc}
	2 &	0 & 0 \\
	0 &-1 & 0 \\
	0 & 0 &-1 \\
\end{array} \right).
\label{4.18}
\end{eqnarray}
We see that $\epsilon_i^{(*)}$ are eigenvectors of $I_3$ and $Y$ under their adjoint action 
corresponding to the (first generation) leptons and to the $W^\pm$ bosons: 
\begin{eqnarray}
&&[I_3,\epsilon_1^*]=\epsilon_1^*, \quad [Y,\epsilon_1^*]=0, \quad \epsilon_1^*\leftrightarrow W^+  \nonumber \\
&&[I_3,\epsilon_2^*]=-\frac{1}{2}\epsilon_2^*, \qquad [Y,\epsilon_2^*]=-\epsilon_2^*\quad \epsilon_2^*\leftrightarrow e_L^-\nonumber \\
&&[I_3,\epsilon_3^*]=\frac{1}{2}\epsilon_3^*, \qquad [Y,\epsilon_3^*]=-\epsilon_3^*\Rightarrow \epsilon_3^*\leftrightarrow\nu_L  \nonumber \\
&&(\epsilon_1\leftrightarrow W^-,\quad \epsilon_2\leftrightarrow e_R^+,\quad
\epsilon_3\leftrightarrow \overline{\nu}_R).
\label{4.19}
\end{eqnarray}
The form of $I_3$ and $Y$ depends on the generation, - i.e., on the choice of  $E_i$ which commutes with the corresponding subalgebra    
$J^8_2 \subset  J^8_3$. We have:
\begin{eqnarray}
2I_3^{(i)}&=&[\epsilon^*_i,\epsilon_i], \quad 3Y^{(i)}=[\epsilon^*_{i-1},\epsilon_{i-1}]-[\epsilon^*_{i+1},\epsilon_{i+1}] \nonumber \\
&&\textit{for} \quad i=1,2,3 \ (mod3). 
\label{4.20}
\end{eqnarray}
Correspondingly, the $W^{\pm}$ boson for the i-th generation is associated with $\epsilon_i^{(*)}$. The generations are 
permuted by the element
\begin{eqnarray}
 \tau=\epsilon_1+\epsilon_2+\epsilon_3 \Rightarrow \tau \epsilon_i^{(*)}\tau^*=\epsilon_{i+1(mod3)}^{(*)},\quad \tau^3=1.
\label{4.21}
\end{eqnarray}
(where $\epsilon_i^{(*)}$ stands for either $\epsilon_i$ or $\epsilon_i^*)$.
Coming back to the first generation, we note that the operators $I_3$ and  $Y$ act on the part $X(\xi,a)$ (\ref{3.27}) of $\mathcal{J}$ by commutation and on the matrix $Z$ (\ref{3.28}) by right multiplication with $(-I_3,-Y)$. Their eigenvectors are given by
\begin{eqnarray}
	Z_1&=&(z^j_1,0,0), \qquad Z_2=(0,z^j_2,0), \qquad Z_3=(0,0,z^j_3)
	\label{4.22}
\end{eqnarray}
\begin{eqnarray}
&& Z_1(-I_3)=0, \quad  Z_2(-I_3)=-\frac{1}{2}Z_2, \quad Z_3(-I_3)=\frac{1}{2}Z_3   \nonumber \\
&& Z_1(-Y)=-\frac{2}{3}Z_1 \quad Z_{2,3}(-Y)=\frac{1}{3}Z_{2,3},
\label{4.23}
\end{eqnarray}	
 while their conjugates have the same eigenvalues with opposite sign and the diagonal matrices $X(\xi,0)$ correspond to $I_3=Y=0$ $(I=0,1)$\footnote{The element $X(\xi,0)=\xi_1\Lambda_3$ carries total isospin $I=1$ while $\xi_2\Lambda_8$ corresponds to $I=0$, both having $I_3=0$.}. Comparing with the quantum numbers of (\ref{4.13}) we end up with the following correspondence
\begin{eqnarray}
	 Z_1\rightarrow d_R, \ Z_2 \rightarrow d_L, \ Z_3 \rightarrow u_L, \ \nonumber \\
	( \overline{Z}_1\rightarrow \overline{d}_L, \ \overline{Z}_2 \rightarrow \overline{d}_R, \ \overline{Z}_3 \rightarrow \overline{u}_R ).
\label{4.24}
\end{eqnarray}
Thus the basic representation \underline{26} of $\rm{F}_4$ splits when restricted to $\rm{SU}(3)_f\times\rm{SU}(3)_c$ into the following irreducible components:
\begin{eqnarray}
	\underline{26}=\underline{\overline{3}}\otimes\underline{3}+\underline{3}
	\otimes\underline{\overline{3}}+\underline{8}\otimes\underline{1}
\label{4.25}
\end{eqnarray}
where the adjoint representation \underline{8} of the flavour  $\rm{SU}(3)=\rm{SU}(3)_f$ consists of two doublets (of leptons and antileptons) and of the three massive gauge bosons and the photon.
\subsection{The symmetry group of the standard model}
It has been observed by Baez and Huerta \cite{BH} that the gauge group of the SM,
\begin{eqnarray}
G_{SM}=	S(U(2)\times U(3))=\frac{SU(2)\times SU(3)\times U(1)}{\mathbb{Z}_6}
\label{4.26}
\end{eqnarray}
can be obtained as the intersection of the Georgi-Glashow and Pati-Salam
\textit{ grand unified theory groups} $SU(5)$ and $(SU(4)\times  SU(2)\times SU(2))/Z_2$ viewed as subgroups of Spin(10).
Here we elaborate on the suggestion of \cite{TD} that one can deduce the symmetry of the SM by applying the Borel - de Siebenthal theory to \textit{admissible} maximal rank subgroups of $\rm{F}_4$ - those that contain the exact colour symmetry $\rm{SU}(3)_c$.
Our analysis in this chapter demonstrates that the intersection of the two admissible maximal subgroups $\rm{Spin}(9)$ and  $\frac{SU(3)\times SU(3)}{\mathbb{Z}_3}$ of $\rm{F}_4$ is precisely the symmetry group $G_{SM}$  (\ref{4.23}) of the SM. The constraints on the $U(1)_Y$ term coming from factoring the $6$-element central subgroup $\mathbb{Z}_6$ imply
\begin{eqnarray}
I_3+\frac{3}{2}Y(=Q+Y)\in\mathbb{Z}
\label{4.27}
\end{eqnarray}
(where $Q=I_3+\frac{1}{2}Y$ is the electric charge); furthermore, if the central element of  $\rm{SU}(3)_c$ is represented by a nontrivial eigenvalue
$\omega$, i.e. if
\begin{eqnarray}
\omega^2 +\omega+1=0 \quad \text{then} \quad  U_Y^2+U_Y+1=0 \quad \text{for} \quad U_Y=e^{2\pi iY}.
\label{4.28}
\end{eqnarray}
The "colourless" Cartan elements $I_3$, $Y$ complemented with the total isospin $I$, $(I_1^2+I_2^2+I_3^2=I(I+1))$ and  the chirality $\gamma$ (which has eigenvalue $1$ for left and $-1$ for right chiral fermions) completely characterize all $32$ first generation fermions as well as the four electroweak gauge bosons. The colour index $j$ can be related to the eigenvalues  of the $\rm{SU}(3)_c$ Cartan matrices, but different "colours" are physically indistinguishable. We only need the chirality $\gamma$ to separate $\nu_R (\gamma=-1)$ from  $\overline{\nu}_L(\gamma=1)$.
\smallskip

\section{Extended euclidean Jordan algebra; the 32 primitive idempotents of the fermion states}
\setcounter{equation}{0}
\renewcommand\theequation{\thesection.\arabic{equation}}
The purpose of this section is to relate our treatment of fundamental fermions in Sect. 4 with the recent realization \cite{D-VT} that a natural euclidean extension of the special Jordan algebra $J^8_2=JSpin_9$ plays the role of the quantum observable algebra of the first generation of the SM.
We shall proceed in three steps. First, we recall the notion of an euclidean extension of a special Jordan algebra \cite{D-VT} as applied to $J^8_2$. Secondly, we provide a fermion Fock space construction of the complexified associative envelope  $C\ell_9$ of  $J^8_2$, inspired by a similar construction by Furey \cite{F18}. We construct 32 \textit{primitive idempotents} \cite{A95, T85/16} which diagonalize three commuting superselection rules and span two irreducible Weyl spinor representations of the Lie algebra $so(9,1)$. Finally, we recover the octet of weak doublets of Sect. 4.1 by identifying the weak isospin $I_3$ with the difference $I^L_3-I^R_3$ of the "left" and the "right" Pati-Salam isospins.
\subsection{Euclidean extension of $J^8_2$. Fermion Fock space}
The associative envelope of the spin factor $J_2^8$ is given by the real Clifford algebra  $C\ell_9=\mathbb{R}[16] \oplus \mathbb{R}[16]$  which is again a Jordan algebra (under anticommutation) but is not euclidean. As observed in \cite{D-VT} the Jordan subalgebra of hermitian matrices $\mathcal {H}_{16}(\mathbb{C}) \oplus \mathcal{H}_{16}(\mathbb{C})$ of 
the complexified Clifford algebra $C\ell_9(\mathbb{C})$ is euclidean (and has the same real dimension $2\times 16^2$, as the above real associative
envelope). It provides a \textit{canonical euclidean extension} of the 10-dimensional Jordan algebra $J^8_2$. The splitting of $C\ell_9$ into two real matrix algebras corresponds to the splitting of the spinorial representation of $so(9,1)$ into two irreducible 16 dimensional chiral Majorana-Weyl spinors. Indeed the Lie algebra $so(9,1)$ belongs to the even part $C\ell^0(9,1)$ of the Clifford algebra $C\ell(9,1)$ which is isomorphic to $C\ell_9$. It is instructive to present (following \cite{BH09}) the mapping $\Gamma$ from $J_2^8$ to $C\ell(9,1)$ through its action on $\underline{16}^+\oplus \underline{16}^-$:
\begin{equation}
\Gamma(X)(\psi_+, \psi_-)=(\tilde{X}\psi_-, X\psi_+)
\label{5.1}
\end{equation}
where $\tilde{X}$ is the \textit{trace reversal} $\tilde{X}=X-(trX)1$.

%We shall  display the 32 dimensional Dirac spinor representation of $so(9,1)$ using the fermionic Fock space realization of $C\ell(9,1)$ (see Sect.4.2).To this end we complement the set of four pairs of fermionic creation and annihilation operators 
\begin{eqnarray}
a_{\mu}^*=\Gamma_{\mu}^-,\quad a_{\mu}=\Gamma_{\mu}^+,\quad \mu=0,1,2,4 \quad \Gamma_\mu^\pm=\sigma_1\otimes\hat{P}_\mu^\pm \quad \text{see eq. (\ref{2.40})} 
\nonumber
\end{eqnarray}
with another pair $(a_8^{*},a_8)$ with all five pairs satisfying 
\begin{eqnarray}
a_{\mu}+a_{\mu}^*=\Gamma_{\mu},\quad \mu=0,1,2,4,8, \quad a_0-a_0^*=i\Gamma_7,  \nonumber \\
a_j-a_j^*=\Gamma_{3j(mod7)},\qquad a_8-a_8^*=\Gamma_{-1}
\label{5.2}
\end{eqnarray}
where $\Gamma_{-1} (=\Gamma(1)$ of (\ref{5.1})) is the (antihermitian) 10th generator  of  $C\ell(9,1)$
satisfying $\Gamma_{-1}^2=-1$ ($[\Gamma_{-1},\Gamma_a]_+=0, \quad a=0,1,\dots ,8$). In accord with our preceding conventions, we identify the electric charge $Q$, the hypercharge $Y$ and the charge $B-L$  with the quadratic expressions:
\begin{eqnarray}
Q&=& \frac{1}{3}(a_1^*a_1+a_2^*a_2+a_4^*a_4)-a_8^*a_8 \nonumber\\	
Y&=& \frac{2}{3}(a_1^*a_1+a_2^*a_2+a_4^*a_4)-a_0^*a_0-a_8^*a_8\nonumber \\
B-L&=& \frac{1}{3}\left([a_1^*,a_1]+[a_2^*,a_2]+[a_4^*,a_4]\right). 
\label{5.3}
\end{eqnarray}
\smallskip
	
\subsection{Fermion state space in terms of primitive \\ 
idempotents}
The states  of an euclidean Jordan algebra are given by idempotents - (hermitian) projection operators. The fermion creation and annihilation operators (\ref{5.1}) allow to define five pairs of commuting idempotents
\begin{eqnarray}
\pi_{\mu}=a_{\mu}a_{\mu}^*, \quad \overline{\pi}_{\mu}=a^*_{\mu}a_{\mu}, \quad \pi_{\mu}+\overline{\pi}_{\mu}=1, \quad \mu=0,1,2,4,8.
\label{5.4}
\end{eqnarray}
The canonical anticommutation relations imply the defining relation for idempotents:
\begin{eqnarray}
\pi_{\mu}^2=\pi_{\mu}, \quad (\overline{\pi}_{\mu}^2=\overline{\pi}_{\mu}).
\label{5.5}
\end{eqnarray}
The products of $\pi_{\mu}$ are again idempotents. An idempotent is called \textit{primitive} if it cannot be decomposed into a sum of (nonzero) mutually orthogonal idempotents. Any product of less than all five $\pi_{\mu}$'s is not primitive. For instance, the product of the first four $\pi_{\mu}$ splits into the following two:
$$\pi_0\pi_1\pi_2\pi_4=\pi_0\pi_1\pi_2\pi_4\pi_8+\pi_0\pi_1\pi_2\pi_4\overline{\pi}_8.$$
Examples of primitive idempotents are:
\begin{eqnarray}
\Omega=\pi_0\pi_1\pi_2\pi_4\pi_8, \qquad \overline{\Omega}=\overline{\pi}_0\overline{\pi}_1\overline{\pi}_2\overline{\pi}_4\overline{\pi}_8.
\label{5.6}
\end{eqnarray}
All mixed products of five factors $\pi_{\mu}$, $\overline{\pi}_{\nu}$  are also primitive. Thus, there are exactly $2^5=32$ primitive idempotents  corresponding to the pure states of the 32 fermions of a given generation (in our case, the first). We postulate, following \cite{T85/16}, that the trace of each primitive idempotent is one. Then the trace of each $\pi_{\mu}$ (and each $\overline{\pi}_{\nu}$) is found to be 16 while the trace of the unit operator equals to \textit {the rank, 32 of the extended Jordan algebra}. The chirality 
\begin{eqnarray}
\gamma=[a_0^*,a_0][a_1^*,a_1][a_2^*,a_2][a_4^*,a_4][a_8^*,a_8] 
\label{5.7}
\end{eqnarray}
commutes with the generators $\Gamma_{ab}=\frac{1}{2}[\Gamma_a,\Gamma_b]$, $a,b=-1,0,\dots,8$ of $so(9,1)$ and splits its 32 dimensional spinor representation into two irreducible chiral components. The right chiral representation $\underline{16}^-$ is obtained from the lowest weight state $\Omega$ by acting on it with the ten raising operators $a_{\mu}^*a_{\nu}^* (\mu<\nu)$. The left chiral fermions  $\underline{16}^+$ are obtained by acting on the highest weight vector $\overline{\Omega}$ by the ten lowering operators $a_{\mu}a_{\nu}$. We note that the opeartors $a_{\mu}a_{\nu}$ and their conjugate belong to the real form $so(5,5)$ of the complexified Lie algebra $so(10)$. The 25-dimensional maximal rank subalgebra $u(3,2)$ of $so(5,5)$ with generators $\frac{1}{2}[a_{\mu}^*,a_{\nu}]$ splits each of the chiral 16-plets into three irreducible components:
\begin{eqnarray}
\underline{16}^-&=& \underline{1}_R+\underline{10}_R+\underline{5}_R;\quad
\underline{16}^+= \underline{1}_L+\underline{10}_L+\underline{5}_L, \quad
\underline{1}_R=\Omega, \quad \underline{1}_L=\overline{\Omega}, \nonumber\\
\underline{10}_R&=& Span\{a^*_{\mu}a^*_{\nu}\Omega a_{\nu}a_{\mu}, \ \mu<\nu\},\quad \underline{5}_R = Span\{a_{\mu}\overline{\Omega}a_{\mu}^*,\ \mu=0,1,2,4,8\} \nonumber \\
\underline{10}_L&=& Span\{a_{\mu}a_{\nu}\overline{\Omega} a^*_{\nu}a^*_{\mu},\ \mu<\nu\},\quad \underline{5}_L = Span\{a_{\mu}^*\Omega a_{\mu},\ \mu=0,1,2,4,8\}.\nonumber \\ 
\label{5.8}
\end{eqnarray}
They are familiar from the Georgi-Glashow $SU(5)$ Grand Unified Theory  \cite{BH, W}.

\smallskip

\subsection{Nonabelian superselection sectors}
The leptons are given by conventional (1-dimensional) pure states characterized by the eigenvalues of the \textit{superselected} charges (in the terminology introduced in \cite{WWW}) $(Q,Y,B-L)$   (\ref{5.3}). We have:
\begin{eqnarray}
(Q,Y,B-L)\Omega &=& (0,0,-1)\Omega\Rightarrow \Omega=|\nu_R><\nu_R| \nonumber\\
(Q,Y,B-L)\overline{\Omega} &=& (0,0,1)\overline{\Omega} \Rightarrow \overline{\Omega}=|\overline{\nu}_L><\overline{\nu}_L|; 
\label{5.9}
\end{eqnarray}
similarly
\begin{eqnarray}
a_0^*\Omega a_0&=&|\nu_L><\nu_L|\Leftrightarrow (0,-1,-1) \nonumber\\ a_8^*\Omega a_8&=&|e_L^-><e_L^-|\Leftrightarrow (-1,-1,-1) \nonumber \\
%a_0^*a_8^*\Omega a_8 a_0&=&|e_R^-><e_R^-|\Leftrightarrow (0,-1,-1) %\nonumber\\ a_8^*\Omega a_8&= & |e_L^-><e_L^-|\Leftrightarrow (-1,-1,-1) %\nonumber \\
a_0^*a_8^*\Omega a_8a_0 &=&|e_R^-><e_R^-|=a_1a_2a_4\overline{\Omega}a_4^*a_2^*a_1^* \Leftrightarrow (-1,-2,-1) \nonumber \\
a_0\overline{\Omega} a_0^*&=&|\overline{\nu}_R><\overline{\nu}_R|\Leftrightarrow (0,1,1), \nonumber\\
a_8\Omega a_8^* &=&|e_R^+><e_R^+|\Leftrightarrow (1,1,1) \nonumber \\
a_1^*a_2^*a_4^*\Omega a_4a_2a_1&=&|e_L^+><e_L^+|= 
a_0a_8\overline{\Omega}a_8^* a_0^*\Leftrightarrow (1,2,1).
\label{5.10}
\end{eqnarray}
It is not meaningful to characterize similarly the quark states since the "colour" is not observable (as the non-abelian gauge symmetry $SU(3)_c$ is unbroken). We shall introduce instead $SU(3)_c$-invariant 3-dimensional projection operators which characterize the observable quark states:
\begin{eqnarray}
a_j^*\Omega a_j&=& a_1^*\Omega a_1+a_2^*\Omega a_2+a_4^*\Omega a_4= |\overline{d}_L><\overline{d}_L|\Leftrightarrow \left(\frac{1}{3},\frac{2}{3},-\frac{1}{3}\right), \nonumber\\
 a_0^*a_j^*\Omega a_ja_0&=&|\overline {d}_R><\overline{d}_R|\Leftrightarrow \left(\frac{1}{3},-\frac{1}{3},-\frac{1}{3}\right) \nonumber \\
 a_j^*a_8^*\Omega a_8a_j&=&|\overline {u}_R><\overline{u}_R|\Leftrightarrow \left(-\frac{2}{3},-\frac{1}{3},-\frac{1}{3}\right) \nonumber \\
 a_0^* a_j^*a_8^*\Omega a_8a_ja_0&=&|\overline {u}_L><\overline{u}_L|\Leftrightarrow \left(-\frac{2}{3},-\frac{4}{3},-\frac{1}{3}\right) \nonumber \\ 
a_j\overline{\Omega} a_j^* &=&|d_R><d_R|\Leftrightarrow \left(-\frac{1}{3},-\frac{2}{3},\frac{1}{3}\right) \nonumber \\
a_0a_j\overline{\Omega} a_j^*a_0^* &=&|d_L><d_L|\Leftrightarrow \left(-\frac{1}{3},\frac{1}{3},\frac{1}{3}\right) \nonumber \\ 
a_ja_8\overline{\Omega} a_8^*a_j^* &=&|u_L><u_L|\Leftrightarrow \left(\frac{2}{3},\frac{1}{3},\frac{1}{3}\right) \nonumber \\ 
a_0a_ja_8\overline{\Omega}a_8^*a_j^*a_0^*&=&|u_R><u_R|\Leftrightarrow \left(\frac{2}{3},\frac{4}{3},\frac{1}{3}\right).
\label{5.11}
\end{eqnarray}
(Summation is understood throughout in the repeated index $j=1,2,4$ - as displayed in the first formula.) We note that the antiparticle to a left chiral quark is a right chiral antiquark and vice versa. All three superselection charges (Q,Y,B-L) change sign going from a particle to its antiparticle.

The passage from the 32-dimensional representation of $so(9,1)$ to the 16-dimensional one $so(9)$, displayed in Sect. 4.1, is similar to the passage from Lorentzian Weyl spinors to the Pauli spinors. To display it we introduce the third components of the "left" and "right" isospin of the Pati-Salam model:
\begin{eqnarray}
2I_3^L=\frac{1}{2}\left([a_0^*,a_0]-[a_8^*,a_8]\right),\quad 2I_3^R=-\frac{1}{2}\left([a_0^*,a_0]+[a_8^*,a_8]\right)
\label{5.12}
\end{eqnarray}
and identify the $I_3$ of Sect. 4.1 with their difference
\begin{eqnarray}
I_3=I_3^L-I_3^R=\frac{1}{2}[a_0^*,a_0].
\label{5.13}
\end{eqnarray}
Identifying in addition the hypercharge in the \underline{16} of $so(9)$ with $B-L$ we reproduce the assignment of quantum numbers to the weakly interacting doublets of Sect. 4.1. 
\smallskip

\section{Concluding remarks}
\setcounter{equation}{0}
\renewcommand\theequation{\thesection.\arabic{equation}}
%If our treatment of composition algebras has been relatively complete and self-contained, our survey of (simple) euclidean Jordan algebras was rather superficial and one-sided. As the omitted properties may play a role in future application of the (automorphism group of the) exceptional Jordan algebra, we shall briefly mention here some of them giving on the way appropriate references. To refresh our memory, we note that the requirement for a Jordan algebra $V$ to be euclidean is equivalent to demanding the existence of positive definite inner product in $V$ such that $<uv, w> = <v, uw>$ , - i.e., such that the operator $L_u$ is hermitean (cf. Gross in \cite{FK}).

The idea that exceptional structures in mathematics should characterize the fundamental constituents of matter has been with us since the Ancient Greeks first contemplated the Platonic solids\footnote{Theaetetus, a contemporary of Plato, gave the first mathematical description of the five regular solids. Plato in the dialog Timaeus (c. 360 B.C.) related the four classical elements, earth, air, water and fire to the four regular solids, cube, octahedron, icosahedron and tetrahedron, respectively, while, according to him "God used the dodecahedron for arranging the constellations on the whole heaven". In the 20th century special structures became part of the excitement with the A-D-E classification that includes the exceptional Lie groups - see \cite{McKay} and references therein.}. The octonions, the elements of the ultimate division algebra, have been linked to the Standard Model of particle physics starting with the paper of G\"unaydin and G\"ursey \cite{GG} which related them to the coloured quarks. Vigorous attempts to implement them in superstring theory \cite{GNORS, CH} remained inconclusive. When the idea of a finite quantum geometry emerged \cite{DKM, CL} (see also the recent contributions \cite{CCS, BF, L18, CCS18} and references therein) it became natural to look for a role of special algebraic structures in such a context. The application of the exceptional Jordan algebra $\mathcal{J}=J_3^8$ to the SM, put forward in Chapters 4 and 5 is a continuation of \cite{TD} connected with the recently introduced notion of an extended special euclidean Jordan algebra \cite{D-VT}. It was iniciated by the paper \cite{DV} of Michel Dubois-Violette (see also \cite{CDD} where differential calculus and the theory of connections on Jordan algebras and Jordan modules is developed).

We find it remarkable that our version of the Borel - de Siebenthal theory \cite{BdS} applied to the automorphism group $F_4$ of
$\mathcal{J}$, yields unambiguously the gauge group $G_{SM}$ of the SM and that the relevant irreducible representation \underline{26} of $F_4$ combines in a single multiplet all quarks' and leptons' doublets and the d-quarks' isotopic singlets with the gauge bosons of the electroweak interactions. On the other hand, the extended observable algerba $\mathcal{H}_{16}(\mathbb{C})\oplus \mathcal{H}_{16}(\mathbb{C})$ of $J_2^8$ gives room to the eight primitive idempotents corresponding to the leptons and antileptons of the SM and to the eight 3-dimensional projectors to the internal state space of quarks and antiquarks of the first generation.

In fact, the exceptional Jordan algebra is intimately related to all exceptional Lie groups (Sect. 3.2 - see also \cite{BS, McC}). It will be interesting to reveal the role of the structure group $E_{6(-26)}$ and the conformal group $E_{7(-25)}$ of $\mathcal{J}$ in the physics of the SM. We intend to return to this problem in future work.

\smallskip

{\footnotesize\noindent {\bf Acknowledgments.} I.T. thanks Michel Dubois-Violette for discussions. Both authors thank Ludmil Hadjiivanov for his participation. I.T. acknowledges the hospitality of the IHES (Bures-sur-Yvette) and  the NCCR SwissMAP (Geneva) where part of this work was done.  S. Drenska's  work has been supported by the Bulgarian National Science Fund, DFNI E02/6.}

\bigskip
\newpage
\section*{Appendix A. The Fano plane of imaginary octonions (\cite{B})}\label{secAA}

$$
\includegraphics[width=4cm]{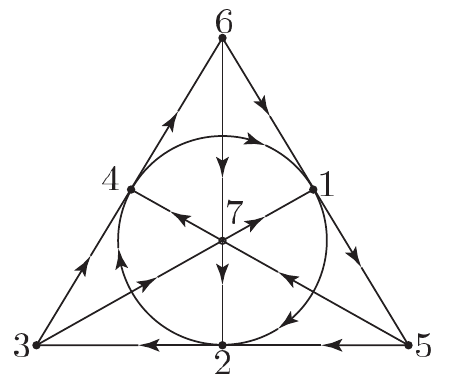}
$$
$$
e_1 = (0,0,1) , e_2 = (0,1,0) \Rightarrow e_1 \, e_2 = e_4 = (0,1,1)
$$
$$
e_3 = (1,0,0) \Rightarrow e_2 \, e_3 = e_5 = (1,1,0)
$$
$$
e_1 \, e_5 = e_6 = (1,1,1)
$$
$$
e_4 \, e_5 = e_7 = (1,0,1) \, .
$$

\centerline{\bf Figure 2.}

\smallskip

\centerline{
	\noindent Projective plane in ${\mathbb Z}_2^3$ with seven points and seven  lines.
}
\bigskip

The multiplication rules for the seven imaginary quaternionic units can be summarized by
$$e_ae_b=-\delta_{ab}+f_{abc}e_c \eqno   ({\rm A}.1) $$
where $f_{abc}$ are fully  antisymmetric and
$$f_{124}=f_{235}=f_{346}=f_{561}=f_{672}=f_{713}=1. \eqno   ({\rm A}.2) $$
The relation  ({\rm A}.2) obey the rules
$$f_{ijk}=1\Rightarrow f_{i+1j+1k+1}=1=f_{2i2j2k}\eqno   ({\rm A}.3) $$
where indices are counted $\mod7$. Eqs.({\rm A}.2) can be recovered from anyone of them and the first relation Eqs.({\rm A}.3).

 We have displayed on Figure 2. the points $e_i$ as non-zero triples of homogeneous coordinates taking values $0$ and $1$ such that the product $e_i \, e_j$ (in clockwise order) is obtained by adding the coordinates $(a,b,c)$, $a,b,c \in \{0,1\}$, modulo two.

%\newpage
\section*{Appendix B. Two bases of $so(8)$ related by the outer automorphism $\pi$.}
\setcounter{equation}{0}
\renewcommand\theequation{\thesection.\arabic{equation}}

The generators $G_{ab}$ of $so(8)$ are given directly by their action on the octonion units (\ref{2.24}):
%\begin{eqnarray}
$$G_{ab}e_b=e_a, \quad G_{ab}e_a=-e_b,\quad G_{ab}e_c=0 \quad \text{for} \quad a\neq c\neq b.
\eqno ({\rm B}.1) $$
%\end{eqnarray}

The action of $F_{ab}$ can also be defduced from definition (\ref{2.24})
and multiplication rules:
%\begin{eqnarray}
$$F_{ab}e_b=\frac{1}{2}e_a, \quad F_{ab}e_a=-\frac{1}{2}e_b, \quad (a \neq b)  F_{ab}=-F_{ba} .$$
$$F_{j0}e_{2j}=\frac{1}{2}e_{4jmod7} (=-F_{0j}e_{2j}) \quad \text{for} \ j=1,2,4, $$
$$ F_{j0}e_{3j}=\frac{1}{2}e_7, \quad F_{70}e_7=\frac{1}{2}e_{3j}, \quad  F_{07}=\frac{1}{2}e_j ,$$
$$F_{0j}e_{6j}=\frac{1}{2}e_{5j}, \quad  F_{j0}e_{5j}=\frac{1}{2}e_{6j}, \quad   \left [F_{j0},F_{0k}\right]=F_{jk}.\eqno   ({\rm B}.2) $$
%\label{A.2}
%\end{eqnarray}
(All indices are counted $mod7$.) From (B.1) and (B.2) we find
%\begin{eqnarray}
$$2F_{0j}=G_{0j}+G_{2j4j}+G_{3j7}+G_{5j6j}$$
$$2F_{03j}=G_{03j}-G_{J7}-G_{2j5j}+G_{4j6j}, \ j=1,2,4 $$
$$2F_{07}=G_{07}+G_{13}+G_{26}+G_{45}.\eqno  ({\rm B}.3) $$
%\label{A.3}
%\end{eqnarray}
In particular, taking the skew symmetry of $G_{ab}$ and the counting $mod7$ into acount we can write
%\begin{eqnarray}
  	$$2F_{02}= G_{02}-G_{14}+G_{35}-G_{76},$$
    $$2F_{04}= G_{04}+G_{12}-G_{36}-G_{75},$$
	$$2F_{03}= G_{03}-G_{17}-G_{25}+G_{46},$$
	$$2F_{06}= G_{06}+G_{15}-G_{27}-G_{43},$$
	$$2F_{05}= G_{05}-G_{16}+G_{23}-G_{47}. \eqno  ({\rm B}.4) $$
%\end{eqnarray}
Note that with the $abc$ are ordering $(1,2,4,3,7,5,6)$ The first (positive) indices of $G$ ($2,3,5$; $1,3,7$; $1,2,4$) correspond to quaternionic triples:$e_2e_3=e_5$, $e_1e_3=e_7$, $e_1e_2=e_4$. Setting

\begin{eqnarray}
G_1&=& \left( \begin{array}{c}
G_{01} \\
G_{24} \\
G_{37} \\
G_{56} \\
\end{array} \right),
G_2= \left( \begin{array}{c}
G_{02} \\
G_{14} \\
G_{35} \\
G_{76} \\
\end{array} \right),
G_4= \left( \begin{array}{c}
G_{04} \\
G_{12} \\
G_{36} \\
G_{75} \\
\end{array} \right),
G_3= \left( \begin{array}{c}
G_{03} \\
G_{17} \\
G_{25} \\
G_{46} \\
\end{array} \right), \nonumber \\
G_7&= &\left( \begin{array}{c}
G_{07} \\
G_{13} \\
G_{26} \\
G_{45} \\
\end{array} \right),
 G_5= \left( \begin{array}{c}
G_{05} \\
G_{16} \\
G_{23} \\
G_{47} \\
\end{array} \right),
G_6= \left( \begin{array}{c}
G_{06} \\
G_{15} \\
G_{27} \\
G_{43} \\
\end{array} \right),
\nonumber
\end{eqnarray}
and similarly for $F_1, \dot{...}F_6$ we find
$$F_a=X_aG_a, \ a=1,2,4,3,7,5,6, \text{with}$$
\begin{eqnarray}
 X_1&=& X_7= \frac{1}{2}\left( \begin{array}{cccc}
1 &  1 &  1 &  1 \\
1 &  1 & -1 & -1 \\
1 & -1 &  1 & -1 \\
1 & -1 & -1 &  1 \\
\end{array} \right),
X_2=X_5=\frac{1}{2} \left( \begin{array}{cccc}
1 & -1 &  1 & -1 \\
-1 &  1 &  1 & -1 \\
1 &  1 &  1 &  1 \\
-1 & -1 &  1 &  1 \\
\end{array} \right), \nonumber \\
X_4&=&X_6=\frac{1}{2} \left( \begin{array}{cccc}
1 &  1 & -1 & -1 \\
1 &  1 &  1 &  1 \\
-1 &  1 &  1 & -1 \\
-1 &  1 & -1 &  1 \\
\end{array} \right),
X_3= \frac{1}{2}\left( \begin{array}{cccc}
1 & -1 & -1 &  1 \\
-1 &  1 & -1 &  1 \\
-1 & -1 &  1 &  1 \\
1 &  1 &  1 &  1 \\
\end{array} \right).
\nonumber
\end{eqnarray}
They all define involutive transformations:
%\begin{eqnarray}
$$X_k^2=\un,\ \det X_k=-1,\ k=1,2,3,4. \eqno   ({\rm B}.5) $$
and satisfy
\begin{eqnarray}
&X_1X_2&=\left( \begin{array}{cccc}
		0 &  0 &  1 &  0 \\
		0 &  0 &  0 & -1 \\
		1 &  0 &  0 &  0 \\
		0 & -1 &  0 &  0 \\
	\end{array} \right)=-X_3X_4,
	X_1X_3=\left( \begin{array}{cccc}
		0 &  0 &   0 &  1 \\
		0 &  0 &  -1 & 0 \\
		0 & -1 &  0 &  0 \\
		1 &  0 &  0 &  0 \\
	\end{array} \right)=-X_2X_4 \nonumber \\
&X_1X_4&=\left( \begin{array}{cccc}
		0 &  1 &  0 &  0 \\
		1 &  0 &  0 &  0 \\
		0 &  0 &  0 & -1 \\
	    0 &  0 & -1 &  0 \\
	\end{array} \right)\qquad \Rightarrow \qquad
	X_1X_2X_3X_4=-\un. \nonumber \\
	&(X_iX_j&=X_jX_i,  \qquad X_1X_2+X_3X_4=0,\qquad X_1X_3+X_2X_4=0 ).
	\qquad	 ({\rm B}.6)
	\nonumber	
\end{eqnarray}
%\label{A.6}
%\end{eqnarray}
%\end{\appendix}
 	
\end{document}